\documentstyle[prd,aps,eqsecnum,amsfonts]{revtex}
\draft
\begin{document}
\newcommand{\artanh}{{\rm artanh}\,}
\newcommand{\Robs}{{\cal R}}
\newcommand{\eps}{\epsilon}
\newcommand{\spur}{{\rm Sp}\,}
\newcommand{\M}{{\cal M}}
\newcommand{\bra}{\langle}
\newcommand{\ket}{\rangle}
\newcommand{\nl}{\displaystyle}
\newcommand{\EL}{\\ \displaystyle}
\newcommand{\HH}{{\cal H}}
\newcommand{\A}{{\cal A}}
\newcommand{\W}{{\omega}}
\newcommand{\then}{\Longrightarrow}
\newcommand{\spr}[2]{\sigma(#1,\, #2)}
\newcommand{\tbf}[1]{{\bf #1}}
\newcommand{\tit}[1]{{\it #1}}
\newcommand{\cl}[1]{{\cal #1}}
\newcommand{\half}{{\scriptstyle \frac12}}
\newcommand{\OPS}{\HH^{(1)}}
\newcommand{\F}{{\cal F}}
\newcommand{\KS}{{\Sigma}}
\newcommand{\vect}[1]{\left(\begin{array}{c}#1\end{array}\right)}
\newcommand{\mat}[1]{\left(\begin{array}{cc}#1\end{array}\right)}
\newcommand{\Rcite}[1]{ref.\cite{#1}}
\newcommand{\Alg}{{\cal U}}
\newcommand{\RWA}{{\cal U}_R}
\newcommand{\LWA}{{\cal U}_L}
\newcommand{\DWA}{\tilde{\cal U}}
\newcommand{\intRR}{\int\limits_{-\infty}^{\infty}}
\newcommand{\intR}{\int\limits_{0}^{\infty}}
\newcommand{\KMS}{{\rm KMS}}
\newcommand{\WF}{{\cal W}}
\newcommand{\LRD}[1]{\frac{{{\displaystyle\leftrightarrow}
\atop {\displaystyle\partial}}}{\partial #1}}
\newcommand{\lrd}[1]{\stackrel{\displaystyle \leftrightarrow}
{\displaystyle\partial_{#1}}}
\newcommand{\diff}[1]{\partial /{\partial #1}}
\newcommand{\Diff}[1]{\frac{\partial}{\partial #1}}
\newcommand{\DiffM}[2]{\frac{\partial #1}{\partial #2}}
\newcommand{\DiffT}[1]{\frac{\partial^2}{\partial {#1}^2}}
\newcommand{\hc}[1]{#1^{\dagger}}
\newcommand{\Ao}{{\rm a}}
\newcount\secnum \secnum=0
\def\newsec{\advance\secnum by 1
\vskip 0.5cm
{\tbf{\the\secnum.\quad}}\nobreak}


\title{Boundary conditions in the Unruh problem}
\author{
N.B. Narozhny\thanks{E-mail: narozhny@pc1k32.mephi.ru},
A.M. Fedotov\thanks{E-mail: a\_fedotov@yahoo.com},
B.M. Karnakov,
V.D. Mur,}
\address{Moscow State Engineering Physics Institute,
115409 Moscow, Russia}
\author{and
V.A. Belinskii\thanks{E-mail: volodia@vxrmg9.icra.it}}
\address{INFN and ICRA, Rome University "La Sapienza",00185 Rome, Italy}

\maketitle
\begin{abstract}
We have analyzed the Unruh problem in the frame of quantum field
theory and have shown that the Unruh quantization scheme is valid
in the double Rindler wedge rather than in Minkowski spacetime.
The double Rindler wedge is composed of two disjoint regions ($R$-
and $L$-wedges of Minkowski spacetime) which are causally
separated from each other. Moreover the Unruh construction implies
existence of boundary condition at the common edge of $R$- and
$L$-wedges in Minkowski spacetime. Such boundary condition may be
interpreted as a topological obstacle which gives rise to a
superselection rule prohibiting any correlations between $r$- and
$l$- Unruh particles. Thus the part of the field from the
$L$-wedge in no way can influence a Rindler observer living in the
$R$-wedge and therefore elimination of the invisible "left"
degrees of freedom will take no effect for him. Hence averaging
over states of the field in one wedge can not lead to
thermalization of the state in the other. This result is proved
both in the standard and algebraic formulations of quantum field
theory and we conclude that principles of quantum field theory
does not give any grounds for existence of the "Unruh effect".
\end{abstract}



\section{Introduction}

It was proposed more than 20 years ago that a detector moving with
constant proper acceleration in empty Minkowski spacetime (MS)
responses as if it had been immersed into thermal bath of Fulling
particles at Davies - Unruh temperature \cite{Unruh,Dav}
\begin{equation}
\label{DUT}
T_{DU}=\frac{\hbar g}{2\pi c k_B},
\end{equation}
where $g$ is proper acceleration of the observer and $k_B$ is
Boltzmann constant. Moreover it is claimed that such response is
universal in the sense that it is the same for any kind of the
detector. This statement is now referred as the "Unruh effect",
see e.g.
Refs.\cite{Isr,Dow,Scia,BD,Sew,GMR,Starob,Tak,GF,GMM,Wald} and
citation therein. More precisely the Unruh effect means that from
the point of view of a uniformly accelerated observer the usual
vacuum in MS occurs to be a mixed state described by thermal
density matrix with effective temperature (\ref{DUT}). In this
paper we will give a critical analysis of this statement and will
show that  fundamental principles of quantum field theory do not
give any physical grounds to assert that the Unruh effect exists.

In fact there are two aspects of the Unruh problem: (i) behaviour
of a particular accelerated detector and (ii) interpretation of
properties of quantum field restricted to a subregion of MS. The
second aspect seems to be more fundamental since one doesn't need
to consider the structure of the detector and details of its
interaction with the quantum field. Indeed the original derivation
of the Unruh effect \cite{Unruh} (see also the later publications
\cite{Isr,Dow,Scia,BD,Sew}) is based only on quantum field theory
principles and use some special models of detectors only as
illustration. Moreover exactly this approach gives grounds for the
assertion about the universality of the detector response.

In this paper we will deal basically with the quantum field theory
aspect (ii) of the Unruh problem. It should be emphasized that
this aspect of the problem is of general interest for quantum
field theory. There are serious arguments (see
e.g.\cite{Unruh,Isr,Dow,Scia,BD,Sew,GMR,Starob,Tak,GF,GMM,Wald})
to think that the Unruh effect is closely related to the effect of
quantum evaporation of black holes predicted by Hawking
\cite{Hawk}. It is claimed that both effects arise due to the
presence of event horizons and that Schwarzschild observer in
Kruskal spacetime may be considered by analogy with Rindler
observer in MS. Furthermore very recently there were proposed some
arguments that evaporation of an eternal Schwarzschild black hole
may be considered as Unruh effect in the six dimensional embedding
MS \cite{6Dim1,6Dim}.

The standard explanation of the Unruh effect is based on existence
of the aforementioned event horizons $h_{\pm}$ bordering the part
of MS which is accessible for a Rindler observer, the
so-called $R$- wedge, see Fig.\ref{Fig2}. In the context of (ii)
aspect of the Unruh problem we refer the term "Rindler observer"
to a uniformly accelerated point object whose trajectory is
entirely located in the $R$-wedge so that the totality of world
lines of all Rindler observers completely cover the interior of
the $R$-wedge. \footnote{For definiteness we assume that Rindler
observers are moving in $z$- direction with respect to some
inertial observer.}

Since such observer due to the presence of horizons have access
only to a part of information possessed by inertial observers it
is commonly accepted that he sees the usual vacuum state in MS as
a mixed state. To put this idea on precise grounds Unruh suggested
\cite{Unruh} a new quantization scheme for a free field in MS
alternative to the standard one. There are two sorts of particles
in this scheme, namely $r$- particles living everywhere but in
$L$- wedge and $l$-particles living everywhere but in $R$- wedge.
$r$-particles as seen by a Rindler observer turn out to be nothing
but the Fulling particles \cite{Fulling}. The corresponding modes
carry only positive frequencies with respect to $(t,z)$- plain
Lorentz boost generator. The parameter of Lorentz boost may be
chosen as time variable in the interior of the $R$-wedge and is
called Rindler time. $r$- and $l$-particle content of Minkowski
vacuum in the Unruh construction can be found by some formal
manipulations \cite{Unruh} and after elimination of non-visible
for a Rindler observer degrees of freedom corresponding to $l$-
particles the "thermal" density matrix with temperature
(\ref{DUT}) can be obtained, see e.g. Refs.\cite
{Scia,GF,GMM,Wald}.

Such construction is however inconsistent because the Unruh
quantization is unitary inequivalent to the standard one
associated with Minkowski vacuum. Therefore the aforementioned
expression for "Minkowski vacuum" in terms of $r$- and $l$-
particles content doesn't make direct mathematical sense and the
"thermal" density matrix which arises after elimination of $l$-
particles degrees of freedom actually vanishes. This difficulty
was pointed out in literature by many authors (see e.g.
Refs.\cite{Sew,WK,Wald}).

Therefore mathematically more rigorous methods based on algebraic
approach to quantum field theory were applied to the problem. In
the frame of this approach the notion of Kubo-Martin-Schwinger
(KMS) state \cite{HHW,Haag} is usually used instead of thermal
equilibrium state (which can not be rigorously defined in this
problem). Reformulation of the Unruh construction on the language
of algebraic approach was presented in Ref.\cite{Kay}. It is worth
to note that mathematical physicists commonly identify
\cite{Sew,FR,Haag} the Unruh effect and the so-called Bisogniano-
Wichmann theorem \cite{BW,BW1}. This theorem is equivalent to the
statement that the Minkowski vacuum state (understood in algebraic
sense) when restricted to the wedge $R$ of MS satisfies the KMS
condition with respect to Rindler time and the corresponding
"temperature" parameter after evaluation in terms of the observer
proper time is exactly the same as given by Eq.(\ref{DUT}).

But apart from the difficulties related to unitary inequivalence
between the Unruh and standard quantization schemes there are also
inconsistencies in physical interpretation of the Unruh
construction which (unlike mathematical difficulties) can not be
resolved. Indeed it is well known that before any measurement
could be carried out one should have prepared the initial state of
the quantum system, the Minkowski vacuum state in our case.
However only a part of MS consisting of the interior of the
$R$-wedge is accessible for a Rindler observer. We will refer to
the interior of $R$- or $L$- wedges as Rindler spacetimes (RS).
These spacetimes are separated from the rest of MS by the event
horizons. As a consequence any well defined scheme of quantization
in RS should imply that the quantum field satisfies boundary
condition at the edge $h_{0}$ of RS. In fact this condition is
nothing but the usual requirement of vanishing of the field at
spatial infinity.

Certainly the horizons $h_{\pm}$ arise due to excessive
idealization of the problem. It is evident that in any physical
situation the only thing one can achieve is to accelerate the
detector during arbitrary long but finite period of time. In the
latter case no horizons arise at all. But in this case one deals
with the (i) aspect of the Unruh problem which has nothing to do
with quantization of field in RS and hence with the notion of
Fulling particles.

In virtue of boundary conditions the Unruh quantization actually
can be performed only in the so-called double Rindler wedge rather
than in MS. The former is a disjoint union of the interiors of the
wedges $R$ and $L$ separated causally and in addition by a sort of
"topological obstacle" . The role of the topological obstacle is
played by the boundary condition necessary for the notion of
Fulling particles and absent in the real Minkowski spacetime. It
follows then that Rindler observers have nothing in common with
the field in MS and that Minkowski vacuum can not be prepared by
any manipulations in double RS. Moreover since the left and right
RS are separated state vectors of the field in double Rindler
wedge are represented by tensor products of state vectors
describing the fields from $R$- and $L$- Rindler spacetimes. Since
the fields in these spacetimes \tit {eternally} don't have any
influence on each other only such states are physically realizable
which do not have any correlations between $r$ and $l$ particles.
In other words there is a "superselection rule" acting in the
Hilbert space of states representing quantum fields in the double
Rindler wedge. Let us stress that this sort of "superselection
rule" arises due to eternal absence of interaction between the $R$
and $L$ parts of the field. Therefore Minkowski vacuum state which
is formally represented in the Unruh construction \cite{Unruh} as
a "superposition" of states with different number of $r$ -
particles and the same number of $l$ - particles is physically
unrealizable for the quantum field in the double RS and discussion
of it's thermal or other properties becomes meaningless.

Since the difficulties with interpretation of the Unruh
construction are of physical nature they of course arise also when
one interprets the results obtained in algebraic approach. In
Ref.\cite{Kay} the notion of KMS state was related to the usual
notion of the bath of heated up $r$- and $l$- particles. But it
occurs that these states may be mathematically well-defined only
for the observables which vanish at some neighborhood of the
common edge of the right and left RS. It is nothing but the
mentioned above boundary condition which leads to loss of any
connection between a Rindler observer and MS.

Elucidation of the role of boundary conditions in the Unruh
problem is the central point of this paper. Therefore we begin in
Section \ref{Minkowski} with brief consideration of boundary
conditions at spatial infinity for a free field quantized in a
plain-wave basis in MS. The existence of boundary conditions is
always implied but their discussion is usually omitted in
text-books. Nevertheless disregarding of this point may lead to
mistakes in treating some delicate problems as it happened in our
opinion with the Unruh problem. In Section \ref{Rindler} we
consider boundary conditions for quantum field in RS. This case
slightly differs in some technical details from the case of a free
field in MS due to the absence of mass gap for Fulling particles.
Section \ref{BoostModesSec} is devoted to consideration of
quantization of a free field in MS in the basis of "boost modes".
This scheme of quantization is unitary equivalent to the usual
plain wave quantization and is exploited in the Unruh
construction. We consider the Unruh construction in Section
\ref{UnruhQuant}. We show that the Unruh quantization scheme is
valid only for the double Rindler wedge and can not be used for
derivation of "thermal properties" of Minkowski vacuum with
respect to a Rindler observer. Algebraic approach to the Unruh
problem is discussed in Section \ref{AlgebraicAppr}. Our results
are summarized in Section \ref{Conclusions}.

In Appendix \ref{CALCULATING_B} we present some technical details
of derivation of the expression for boost mode annihilation
operator in terms of the field values on a Cauchy surface. The
derived formula allows one to understand difference between the
Unruh and Fulling operators. Appendix \ref{inequiv} includes
discussion of analogy between the Unruh construction and the
construction of squeezed states for a two dimensional harmonic
oscillator. In this Appendix we have also included a proof of
unitary inequivalence of the standard plain wave and Unruh schemes
of quantization. Although this issue seems to be a well-known fact
we actually could not find a detailed discussion of it in physical
literature.

In the paper we restrict our discussion to the case of massive
neutral scalar field in $1+1$ dimensional MS. Generalization to
higher dimensions may be obtained straightforwardly by introducing
components of momentum $\vec{q}$ orthogonal to the direction of
motion of a Rindler observer just changing definition of the mass
of the field as $m\to\sqrt{m^2+q^2}$ and inserting additional
integration over $\vec{q}$ in appropriate places (see e.g.
Ref.\cite{GMM} for $1+3$ dimensional case and Ref.\cite{Tak} for
the general case of $1+n$ dimensions).

Short presentation of our results can be found in Refs.
\cite{JETP,PhLet}.


\section{Quantization of a neutral scalar field in D=1+1
Minkowski spacetime (plane wave modes)} \label{Minkowski}

In this section we will discuss the boundary conditions for a
quantized free scalar field in two-dimensional Minkowski
space-time (MS). Let
\begin{equation}
\label{Coords_M}
x=\{t,z\},\quad ds^2=dt^2-dz^2,
\end{equation}
be global coordinates and metric of pseudoeuclidean plain
\footnote{From this place we use natural units $\hbar=c=k_B=1$
throughout the paper.}.
Operator of a free neutral scalar field $\phi_M(x)$ of
mass $m$ satisfies the Klein-Fock-Gordon (KFG) equation
\begin{equation}
\label{KFG_M}
\left\{\frac{\partial^2}{\partial t^2}+{\cal
K}_M(z) \right\}\phi_M(x)=0,\quad {\cal
K}_M(z)=-\frac{\partial^2}{\partial z^2}+m^2.
\end{equation}
The plain waves
\begin{equation}
\label{PlainWaves} \Theta_p(x)=(2\epsilon_p)^{-1/2}
e^{-i\epsilon_p t} \varphi_p(z),\quad \varphi_p(z)=(2\pi)^{-1/2}
e^{ipz},\quad \epsilon_p=\sqrt{p^2+m^2},\quad -\infty<p<\infty,
\end{equation}
form a complete set of solutions of the equation (\ref{KFG_M})
orthonormalized relative to scalar product in MS
\begin{equation}
\label{ScalProd_M}
(f,g)_M\equiv i\int\limits_{-\infty}^{\infty}\,
f^*(x)\LRD{t} g(x) dz.
\end{equation}

The completeness of the set (\ref{PlainWaves}) allows one to
perform quantization by setting
\begin{equation}
\label{Quant_PW} \phi_M(x)=\int\limits_{-\infty}^{\infty} dp\,
\left[a_{p}\Theta_{p}(x)+a_{p}^{\dag}\Theta_{p}^{*}(x)\right],
\end{equation}
where $a_{p}$ and $a_{p}^{\dag}$ are annihilation and creation
operators obeying canonical commutation relations. The vacuum
state $|0\rangle_{M}$ in MS is defined by the relations
\begin{equation}
\label{Vac_M}
a_{p}|0_{M}\rangle=0,\quad -\infty<p<\infty,
\end{equation}
and operators $a_{p}$ can be expressed in terms of field operator
$\phi_M(x)$ values on an arbitrary spacelike surface by
$a_{p}=(\Theta_{p},\phi_M)_M$. In particular
\begin{equation}
\label{FieldTo_a} a_{p}=i\int\limits_{-\infty}^{\infty}\,
\Theta_{p}^*(x)\LRD{t}\phi_M(x)\,dz.
\end{equation}

It is commonly assumed that the field $\phi_M(x)$ vanishes at
spatial infinity. Nevertheless it is worth emphasizing that the
operators $a_{p}$, $a_{p}^{\dag}$, $\phi_M(x)$ are unbounded ones
and therefore the requirement $\phi_M(x)\to 0,\;z\to\pm\infty$ as
well as relations (\ref{Quant_PW}),(\ref{FieldTo_a}) should be
understood in weak sense \cite{Wtm,Jst,ReS}. The latter means that
these statements relate to arbitrary matrix elements of operators
under discussion.

Note that the requirement of vanishing of the field $\phi_M(x)$ in
the weak sense at spatial infinity is a necessary condition for
finiteness of the energy of the field. For illustration of this
statement we will consider one particle amplitude
\begin{equation}
\label{OPA}
\phi_{f}(x)=\langle 0_{M} |\phi_M(x)|f\rangle,
\end{equation}
which determines all matrix elements of the free field operator. One
particle state $|f\rangle$ in Eq.(\ref{OPA}) is defined by
\begin{equation}
\label{OPS} |f\rangle=a^{\dag}(f)|0_{M}\rangle,\quad a
^{\dag}(f)=\int\limits_{-\infty}^{\infty} dp\, f(p)
a_{p}^{\dag},\quad \langle
f|f\rangle=\int\limits_{-\infty}^{\infty} dp\, |f(p)|^2=1.
\end{equation}

The field Hamiltonian expectation value in the state
$|f\rangle$ is given by the following expression
\begin{equation}
\label{Hamilt_M}
\langle f| H|f\rangle=\langle f|\,
\int\limits_{-\infty}^{\infty} dz\, :T^{00}:\,
| f\rangle= \frac12\,\int\limits_{-\infty}^{\infty} dz\,
\left\{\left\vert\frac{\partial \phi_{f}(x)}{\partial t}
\right\vert^2+\left\vert\frac{\partial \phi_{f}(x)}{\partial z}
\right\vert^2+m^2\left\vert \phi_f(x)\right\vert^2\right\}.
\end{equation}

One can easily see that the finiteness of the field energy
$\langle f|H|f\rangle$ implies
\begin{equation}
\label{SqInt} \int\limits_{-\infty}^{\infty}
|\phi_f(x)|^2\,dz<\infty,\quad \int\limits_{-\infty}^{\infty}
\left|\frac{\partial\phi_f(x)} {\partial z}\right|^2\,dz<\infty,
\end{equation} and
hence leads to continuity of $\phi_{f}(x)$ and its vanishing at
spatial infinity
\begin{equation}
\label{BoundCond}
\phi_{f}(t,z)\to 0,\quad z\to\pm\infty.
\end{equation}
Indeed from the inequality
\[ |\phi_f^2(t,z_2)-\phi_f^2(t,z_1)|\le 2\int\limits_{z_1}^{z_2} dz\,
\left\vert \phi_f(x) \DiffM{\phi_f(x)}{z}\right\vert \le
2\left(\;\int\limits_{z_1}^{z_2} dz\,|\phi_f(x)|^2\cdot
\int\limits_{z_1}^{z_2}
dz\,\left\vert\DiffM{\phi_f(x)}{z}\right\vert^2 \right)^{1/2} \]
and square integrability of $\phi_f(x)$ and
$\partial\phi_f(x)/\partial z$ it is evident that the function
$\phi_f(t,z)$ is continuous and there exists the limit
$\lim\limits_{z\to\pm\infty}\phi_f(t,z)$. But if $\phi_f(x)$ is
square integrable then this limit should be zero (for details of
the proof in more general case see Sec. 5.6 in Ref.\cite{Rich}).
Hence if we refuse from the condition (\ref{BoundCond}) then the
energy becomes infinite.

This boundary condition in terms of Whightman functions is
equivalent to vanishing of the two-point function for infinite
spacelike separations. From Eqs.(\ref{Quant_PW}), (\ref{Vac_M})
for positive-frequency function $\Delta^{(+)}(x,m)= i\langle
0_M|\phi_M(x)\phi_M(0) |0_M\rangle$ we obtain \cite{Wtm}
\begin{equation}
\label{DeltaPlus}
\Delta^{(+)}(x,m)=\frac14\times\left\{
\begin{array}{cc} \nl
H_0^{(2)}(m\sqrt{t^2-z^2}), & t>|z|, \EL \frac{2i}{\pi}
K_0(m\sqrt{z^2-t^2}), & |t|<|z|, \EL -H_0^{(1)}(m\sqrt{t^2-z^2}),
& t<-|z|,
\end{array}\right.
\end{equation}
where $H_{\nu}^{(1,2)}$ are Hankel and $K_{\nu}$ - Macdonald
(modified Bessel) functions. Therefore using asymptotic expansions
for $H_{\nu}^{(1,2)}$ and $K_{\nu}$ \cite{BE1} we get for
$|z|\to\infty$, $t=0$
\begin{equation}
\label{DeltaPlusAsympt}
\Delta^{(+)}(x,m)\propto (m|z|)^{-1/2} e^{-m|z|}.
\end{equation}
The two-point commutator function
$\Delta(x-x',m)=i[\phi_M(x),\phi_M(x')]$ reads
\begin{equation}
\label{commut} \Delta(x,m)=\frac14\{{\rm sgn}\,(t-z)+{\rm sgn}\,
(t+z)\}J_0(m\sqrt{t^2-z^2}),
\end{equation}
where $J_{\nu}$ denotes Bessel function, $\theta(\tau)$ is the
Heavicide step function and ${\rm
sgn}\,(\tau)=\theta(\tau)-\theta(-\tau)$. Note that the Cauchy
data of the function $\Delta(x,m)$ on the surface $t=0$ is
\begin{equation}
\label{props_comm}
\Delta(x,m)\vert_{t=0}=0,\quad\left.
\DiffM{\Delta(x,m)}{t}\right\vert_{t=0}=\delta(z),
\end{equation}
in full analogy with the Cauchy data for the Pauli-Jordan function
in four dimensional case \cite{Schweb,IZ}.

\parindent 0.7cm
\section{Quantization of a neutral scalar field in D=1+1
Rindler space}
\label{Rindler}

In this section we will consider quantization of a neutral scalar
field in D=1+1 Rindler space the geometry of which is described by
the metric \cite{Rind}
\begin{equation}
\label{Coords_R} ds^2=\rho^2 d\eta^2-d\rho^2,\quad
-\infty<\eta<+\infty,\quad 0<\rho<\infty.
\end{equation}
This issue plays an important role for the Unruh problem because it
defines the notion of Fulling particles. In the Sec.\ref{SUBA} we
will define the Fulling modes which form a basis for quantization
and introduce the notion of Fulling particles \cite{Fulling}. In
the Sec.\ref{SUBB} we will discuss boundary conditions arising in
the procedure of Fulling quantization.

\subsection{Fulling quantization}
\label{SUBA}

KFG equation in RS takes the form
\begin{equation}
\label{KFG_R}
\left\{\DiffT{\eta}+{\cal K}_R(\rho)\right\}\phi_R(\xi)=0,\quad
{\cal K}_R(\rho)=-\rho\Diff{\rho}\rho\Diff{\rho}+m^2\rho^2, \quad
\xi=\{\eta,\rho\}
\end{equation}
Differential operator ${\cal K}_R$ is a self- adjoint positive
operator in the Hilbert space $L^2_{\sigma}(0,\infty)$ of square
integrable functions with measure $d\sigma(\rho)=d\rho/\rho$ and
inner product
$(\chi,\psi)_{L^2}=\int_0^{\infty}\chi^*(\rho)\psi(\rho)\,d\rho/\rho$.
Functions from the domain of it's definition ${\cal D}({\cal
K}_R)$ obey the conditions
\begin{equation}
\label{domain_KR}
(\psi,\psi)_{L^2}=\intR\frac{d\rho}{\rho}\,
|\psi(\rho)|^2<\infty,\quad \intR\frac{d\rho}{\rho}\,
\left\vert\rho\DiffM{\psi(\rho)}{\rho}\right\vert^2<\infty,
\quad\psi(0)=0,
\end{equation}
where the last restriction is a consequence of the two previous
(compare to Eqs.(\ref{SqInt}), (\ref{BoundCond})). The condition
$\psi(0)=0$ is an automatic or built-in boundary condition, see
Ref.\cite{Rich}. This means that at the point $\rho=0$ we
encounter the case of the Weyl limit-point. From mathematical
point of view it results from the fact that deficiency indices of
the operator ${\cal K}_R$ are equal to $(0,0)$. The physical
meaning of this condition was discussed by Fulling \cite{Fulling}
using the language of wave packets.

The Eigenfunctions of the operator ${\cal K}_R$,
\begin{equation}
\label{Eigen_KR}
\psi_{\mu}(\rho)=\pi^{-1}(2\mu\sinh\pi\mu)^{1/2}K_{i\mu}(m\rho),
\quad {\cal K}_R\psi_{\mu}(\rho)=\mu^2\psi_{\mu}(\rho),
\end{equation}
satisfy the orthogonality and completeness conditions,
\begin{equation}
\label{NormComp_R}
\intR\frac{d\rho}{\rho}\,\psi_{\mu}^*(\rho)\psi_{\mu'}(\rho)=
\delta(\mu-\mu'), \quad
\intR d\mu\,\psi_{\mu}(\rho)\psi_{\mu}^*(\rho')=
\rho\delta(\rho-\rho').
\end{equation}
Note that the functions $\psi_{\mu}(\rho)$ satisfy the third
condition in Eq.(\ref{domain_KR}) in the sense of distributions.

For the solution of the Cauchy problem for Eq.(\ref{KFG_R}) we
have
\begin{equation}
\label{Cauch_sol} \phi_R(\eta,\rho)=e^{-i\eta{\cal
K}_R^{1/2}}\psi(\rho)+ e^{i\eta{\cal K}_R^{1/2}}\psi^*(\rho),
\quad \psi(\rho)=\frac12\phi_R(0,\rho)+\frac{i}2{\cal K}_R^{-1/2}
\Diff{\eta}\phi_R(0,\rho).
\end{equation}
Therefore positive-frequency with respect to timelike variable
$\eta$ modes, Fulling modes \cite{Fulling}, read
\begin{equation}
\label{Full_Modes}
\Phi_{\mu}(\xi)=(2\mu)^{-1/2}\psi_{\mu}(\rho)e^{-i\mu\eta},
\quad\mu>0.
\end{equation}
These modes are orthonormal relative to the inner product in RS,
\begin{equation}
\label{Inner_Prod_R}
(F,G)_R=i\intR\frac{d\rho}{\rho}\,F^*(\xi)\LRD{\eta}G(\xi),
\end{equation}
and together with $\Phi_{\mu}^*$ form a complete set of solutions of
KFG equation (\ref{KFG_R}). Therefore they may be used for
quantizing the field $\phi_R$,
\begin{equation}
\label{Quant_R}
\phi_R(\xi)=\intR d\mu\,\{c_{\mu}\Phi_{\mu}(\xi)+
\hc{c_{\mu}}\Phi_{\mu}^*(\xi)\},
\quad [c_{\mu},\hc{c_{\mu'}}]=\delta(\mu-\mu'),
\quad [c_{\mu},c_{\mu'}]=[\hc{c_{\mu}},\hc{c_{\mu'}}]=0,
\end{equation}
and one can define vacuum state $|0_R\ket$ for Rindler space by the
condition
\begin{equation}
\label{Vac_R}
c_{\mu}|0_R\ket=0,\quad \mu>0.
\end{equation}
The states which are created from this vacuum by operators
$\hc{c_{\mu}}$ correspond to Fulling (or sometimes also called
\cite{GMM} Rindler) particles.

The annihilation operator $c_{\mu}$ may be expressed in terms of
the field $\phi_R$ by
\begin{equation}
\label{c_phi_R}
c_{\mu}=(\Phi_{\mu},\phi_R)_R=\frac{i}{\sqrt{2\mu}}\,
\intR\frac{d\rho}{\rho}\,\psi_{\mu}(\rho) \left.
\left(\DiffM{\phi_R(\xi)}{\eta}-i\mu\phi_R(\xi)\right)
\right\vert_{\eta=0}.
\end{equation}
The secondly quantized operator corresponding to the Killing vector
$i\diff{\eta}$,
\begin{equation}
\label{K}
K=\intR\,\mu\hc{c_{\mu}}c_{\mu}\,d\mu,
\end{equation}
plays a role of Hamiltonian.

With the help of Eqs.(\ref{Quant_R}), (\ref{Vac_R}) one can
calculate the two-point commutator
$D(\eta-\eta',\rho,\rho')=i[\phi_R(\xi),\phi_R(\xi')]$ and the
positive-frequency Whightman function
$D^{(+)}(\eta-\eta',\rho,\rho')=i\bra 0_R|\phi_R(\xi)
\phi_R(\xi')|0_R\ket$ for RS,
\begin{mathletters}
\begin{equation}
\label{WightFunc}
D(\eta-\eta',\rho,\rho')=\frac2{\pi^2}\, \intR
d\mu\,\sinh\pi\mu\sin\mu(\eta-\eta')\,
K_{i\mu}(m\rho)K_{i\mu}(m\rho'),
\end{equation}
\begin{equation}
\label{WightFuncPlus}
D^{(+)}(\eta-\eta',\rho,\rho')=\frac{i}{\pi^2}\, \intR
d\mu\,\sinh\pi\mu\cos\mu(\eta-\eta')\,
K_{i\mu}(m\rho)K_{i\mu}(m\rho') +\frac12
D(\eta-\eta',\rho,\rho').
\end{equation}
\end{mathletters}
Using the relation between Minkowski coordinates $(t,z)$ and
Rindler coordinates $(\eta,\rho)$ in the $R$-wedge of MS,
\begin{equation}
\label{CoordConn}
\eta=\artanh(t/z),\quad \rho=(z^2-t^2)^{1/2},
\end{equation}
one can easily see that the two point commutation functions $D$
and $\Delta$ coincide. In particular for
$\Delta s^2=2\rho\rho'\cosh(\eta-\eta')-\rho^2-\rho'^2>0$ we have
$D(\eta-\eta',\rho,\rho')=\half J_0\left(m (\Delta s^2)^{1/2}\right)$.
This coincidence of two-point commutators means that the local
properties of the quantum fields $\phi_M$ and $\phi_R$ are the same.
Nevertheless global properties of these fields are different due to
different definitions of vacuums $|0_M\ket$ and $|0_R\ket$,
$\Delta^{(+)}\ne D^{(+)}$. Note that singularities of these functions
for coinciding points are the same and cancel when one takes their
difference,
\begin{equation}
\label{WightFuncDif}
\bra 0_M|\phi_M^2(\xi)|0_M\ket- \bra 0_R|\phi_R^2(\xi)|0_R\ket=
\frac1{\pi^2}\intR d\mu\,e^{-\pi\mu} K_{i\mu}^2(m\rho)>0.
\end{equation}

\subsection{Boundary condition}
\label{SUBB}

Let us now discuss the boundary conditions for the field
$\phi_R(\xi)$. As in section \ref{Minkowski} we will consider
one-particle amplitude
\begin{equation}
\label{OPA_R}
\phi_g(\xi)=\bra 0_R|\phi_R(\xi)|g\ket=
\exp(-i\eta{\cal K}_R^{1/2}) \psi_g(\rho).
\end{equation}
(compare to Eq.(\ref{Cauch_sol})), where
\begin{equation}
\label{gCg}
|g\ket=\hc{c}(g)|0_R\ket,\quad \hc{c}(g)=\intR
\frac{d\mu}{\mu^{1/2}}g(\mu)\hc{c_{\mu}}.
\end{equation}
The spatial part $\psi_g$ of the one particle amplitude (\ref{OPA_R})
is expressed in terms of the weight function $g$ as follows,
\begin{equation}
\label{psi_g}
\psi_g(\rho)=\intR d\mu\,G(\mu,\rho),\quad G(\mu,\rho)=
\frac{g(\mu)}{\pi} \left(\frac{\sinh\pi\mu}{\mu}\right)^{1/2}
K_{i\mu}(m\rho).
\end{equation}
For inner product in RS we have
\begin{equation}
\label{KG_to_Ltwo} (\phi_g,\phi_h)_R= 2({\cal K}_R^{1/4}\psi_g,
{\cal K}_R^{1/4}\psi_h)_{L^2},
\end{equation}
and the state $|g\ket$ is normalized by the condition
\begin{equation}
\label{norm_g}
\bra g|g\ket=\intR\frac{d\mu}{\mu}|g(\mu)|^2=
2\intR\frac{d\rho}{\rho}|{\cal K}_R^{1/4}\psi_g|^2=1.
\end{equation}

We will discuss boundary conditions at both boundaries of RS,
namely at the points $\rho=\infty$ and $\rho=0$. It's worth to
note that the point $\rho=0$ may also be considered as spatial
infinity. It becomes evident after the Langer transformation
$u=\ln(m\rho)$ ($-\infty<u<\infty$) mapping the singular point
$\rho=0$ into $-\infty$. After this transformation operator
${\cal K}_R$ takes the form
\begin{equation}
\label{KR}
{\cal K}_R=-\DiffT{u}+V(u),\quad V(u)=m^2e^{2u}.
\end{equation}
Since $V(u)$ is a confining potential at $u\to+\infty$ the boundary
condition
\begin{equation}
\label{BC_infty} \phi_g(\eta,+\infty)=0,
\end{equation}
is obvious and is satisfied not only for the amplitude
(\ref{OPA_R}) but even for Eigenfunctions $\psi_{\mu}(\rho)$ of
operator ${\cal K}_R$ . Therefore we concentrate below on the less
evident case $\rho=0$ or $u=-\infty$.

The requirement of finiteness of the average energy (\ref{K}) in
the state $|g\ket$,
\begin{equation}
\label{FiniteEn} \bra
g|K|g\ket=\frac12\intR\frac{d\rho}{\rho}\left\{
\left\vert\DiffM{\phi_g}{\eta}\right\vert^2+
\rho^2\left\vert\DiffM{\phi_g}{\rho}\right\vert^2+
m^2\rho^2\left\vert\phi_g\right\vert^2\right\}= \intR
d\mu\,|g(\mu)|^2<\infty,
\end{equation}
leads to the restrictions
\begin{equation}
\label{rest_R}
\intR\frac{d\rho}{\rho}\,\left\vert\rho\DiffM{\phi_g}{\rho}
\right\vert^2<\infty,\quad\intR\frac{d\rho}{\rho}\,
\left\vert\rho\phi_g\right\vert^2<\infty.
\end{equation}
But from these restrictions (unlike the case we had in MS) it does
not immediately follow that
\begin{equation}
\label{BC_zero} \phi_g(\eta,0)=0.
\end{equation}
Therefore we need use more delicate procedure to prove the
condition (\ref{BC_zero}).

Let us split the integral in Eq.(\ref{psi_g}) into three parts,
\begin{equation}
\label{split}
\begin{array}{l}\nl
\psi_g(\rho)=I_1(\rho)+I_2(\rho)+I_3(\rho),\EL
I_1=\int\limits_0^{\mu_1}G\,d\mu,\quad
I_2=\int\limits_{\mu_1}^{\mu_2}G\,d\mu,\quad
I_3=\int\limits_{\mu_2}^{\infty}G\,d\mu,
\end{array}
\end{equation}
where $\mu_1$, $\mu_2$ are arbitrary numbers such that $0<\mu_1\ll
1\ll\mu_2<\infty$ and the function $G(\mu,\rho)$ is defined in
Eq.(\ref{psi_g}). After applying the known asymptotic behaviour of
the Macdonald function \cite{BE1} we obtain for $u=\ln(m\rho)<0$,
$|u|\gg1$
\begin{equation}
\label{G_asympt}
G(\mu,\rho)=\frac{g(\mu)}{\sqrt{\pi}\mu}\times
\left\{
\begin{array}{lc}\nl
-\sin(\mu u-\mu\ln2),& \mu\ll 1, \EL\EL \cos(\mu
u-\mu\ln{2}-\arg\Gamma(i\mu)),& \mu\sim 1, \EL\EL
\sin(\mu\ln\mu-\mu u+\mu(\ln{2}-1)+\pi/4),& \mu\gg 1.
\end{array}
\right.
\end{equation}

Let us first proceed with the last two terms in Eq.(\ref{split}).
It follows from the normalization condition Eq.(\ref{norm_g}) and
the evident inequality $|g|\le\half (1+|g|^2)$  that the integral
$\int_{\mu_1}^{\mu_2}|g(\mu)|\,d\mu/\mu$ should converge.
Therefore we may apply the Riemann-Lebesgue lemma to conclude that
$I_2(\rho)$ vanishes when $\rho\to 0$. To estimate $I_3$ we use
the Schwartz inequality and get
\begin{equation}
\label{I3}
|I_3(\rho)|^2\le\frac1{\pi\mu_2}\int\limits_{\mu_2}^{\infty}
|g(\mu)|^2\,d\mu.
\end{equation}
Now using the condition of finiteness of the energy
Eq.(\ref{FiniteEn}) we conclude that $I_3(\rho)$ may be done
arbitrary small by the appropriate (independent of the value of
$\rho$) choice of $\mu_2$. Therefore the sum $I_2(\rho)+I_3(\rho)$
tends to zero when $\rho\to 0$. Note also that for differentiable
functions $g(\mu)$ one has the estimation $I_2+I_3\propto
\ln^{-1}(1/m\rho)$ for $\rho\ll m^{-1}$.

Consider now the first part $I_1(\rho)$ of the integral in
Eq.(\ref{split}). We will discuss first the case of weight
functions $g(\mu)$ which are continuous at the point $\mu=0$. From
Eq.(\ref{norm_g}) it immediately follows that such weight
functions should vanish for $\mu\to 0$, $g(0)=0$.

Let $g(\mu)$ vanish for $\mu\to 0$ as a power of $\mu$,
\begin{equation}
\label{power}
g(\mu)=a\mu^{\alpha},\quad\alpha>0,\quad \mu\to 0.
\end{equation}
Then from Eqs.(\ref{split}), (\ref{G_asympt}) we obtain the
estimation
\begin{equation}
\label{I1_power} I_1(\rho)=\frac{a\Gamma(\alpha)\sin(\pi\alpha/2)}
{\sqrt{\pi}(\ln\frac{2}{m\rho})^{\alpha}},\quad \rho\ll
m^{-1},\quad \alpha\ne 2,4,6,...\quad .
\end{equation}
Thus for this case $I_1(\rho)$ decreases logarithmically when
$\rho\to 0$. We see that for $0<\alpha<1$ the term $I_1(\rho)$
dominates in the sum in Eq.(\ref{split}) for small $\rho$ and thus
the whole integral over $\mu$ is defined by the behaviour of
$g(\mu)$ at small $\mu$ while for $\alpha>1$ generally $\psi_g(\rho)$
does not depend on behaviour of $g(\mu)$ at small $\mu$.
For even values of $\alpha$ the leading term (\ref{I1_power})
vanishes and $I_1(\rho)$ decreases for $\rho\to 0$ even faster.
For example, for the particular weight function
\begin{equation}
\label{example} g_0(\mu)=\frac{a_0\mu(\mu\sinh\pi\mu)^{1/2}}
{\sqrt{\pi}\Gamma^2(\epsilon)}\, |\Gamma(\epsilon+i\mu)|^2,\quad
(\epsilon>0), \quad g_0(\mu)\approx a_0\mu^2,\quad\mu\to 0,
\end{equation}
($a_0$ - normalization constant) by performing Kontorovich - Lebedev
transform \cite{KL,BE} we obtain
$\psi_g(\rho)\propto(m\rho)^{\epsilon}\exp(-m\rho)\propto
\rho^{\epsilon}$ for $\rho\ll m^{-1}$.

If the weight function vanishes for small $\mu$ logarithmically,
\begin{equation}
\label{log}
g(\mu)=b\left(\ln\frac1{\mu}\right)^{-\beta},\quad
\beta>0,\quad \mu\to 0,
\end{equation}
then for small $\rho$ we have
\begin{equation}
\label{I1_log}
I_1(\rho)=\frac{\sqrt{\pi}b}
{2\left(\ln\ln\frac2{m\rho}\right)^{\beta}} \left(1- \frac{\beta
C}{\left(\ln\ln\frac2{m\rho}\right)^{\beta}} +...\right),
\end{equation}
where $C=0.577...$ is Euler constant. This asymptotic representation is
valid in particular for $0<\beta<1/2$ when the normalization
integral (\ref{norm_g}) diverges. Since the faster $g(\mu)$
decreases for $\mu\to 0$ the faster $I_1(\rho)$ tends to zero, we
conclude that the one particle amplitude $\Phi_g(\xi)$ tends to
zero when $\rho\to 0$ for all weight functions which are
continuous at $\mu=0$, satisfy the normalization condition and
correspond to the states with finite energy.

Note that continuity of weight functions $g(\mu)$ is not a
necessary condition for the validity of the condition
Eq.(\ref{BC_zero}). As an example we will consider the weight
function which for small $\mu$ behaves as
\begin{equation}
\label{Rich}
g(\mu)=c\,\sigma^{\gamma}\exp(-d\sigma^{\delta}\sin^2\sigma),
\quad\gamma\ge 0,\quad \sigma=\ln(1/\mu).
\end{equation}
For big values of $\sigma$ this function is localized almost near
the points $\sigma_n=\pi n$ with characteristic width
$\Delta\sigma_n=d^{-1/2}(\pi n)^{-\delta/2}$. Therefore for the
contribution of small $\mu$ to the normalization integral
Eq.(\ref{norm_g}) we have
\begin{equation}
\label{Rich_norm} \int\limits_0^{\mu_1}\frac{d\mu}{\mu}
|g(\mu)|^2=\int\limits_{\sigma_1}^{\infty}
d\sigma\,|g(e^{-\sigma})|^2\propto \sum_n d^{-1/2}(\pi
n)^{2\gamma-\delta/2},
\end{equation}
where $\sigma_1=\ln(\frac{1}{\mu_1})$. Therefore the function
$g(\mu)$ with behaviour at small $\mu$ as given by Eq.(\ref{Rich})
may be normalizable if
\begin{equation}
\label{Rich_restr}
\delta>4\gamma+2
\end{equation}
(the case $\gamma=1/2$, $\delta=6$ is known as Dirichlet-Reymond
example \cite{Whit,Rich}). But let us note that for the function
Eq.(\ref{Rich}) with parameters satisfying Eq.(\ref{Rich_restr})
the integral $\int_0^{\mu_1}|g(\mu)|\,d\mu/\mu=
\int_{\sigma_1}^{\infty}|g(e^{-\sigma})|\,d\sigma$ also converges
and thus in virtue of the Riemann-Lebesgue lemma we conclude that
even in this case we have $I_1(\rho)\to 0$ as $\rho\to 0$. It
looks as if the condition Eq.(\ref{BC_zero}) is valid for
discontinues at $\mu=0$ weight functions as well.

Let us also note that for arbitrary small but finite accuracy of
measuring of energy of Fulling particles the weight functions of
the type Eq.(\ref{Rich}) are physically indistinguishable from
those vanishing for small $\mu$. We can state therefore that the
boundary condition Eq.(\ref{BC_zero}) is satisfied at least for
all physically realizable states $|g\ket$.

It is worth to emphasize that the considered boundary condition
(\ref{BC_zero}) for quantized field can not be reduced to the
boundary condition $\psi(0)=0$ for functions from the domain of
definition of the operator ${\cal K}_R$ because the expansion
(\ref{Quant_R}) is valid for nonvanishing (in weak sense) at the
point $\rho=0 $ functions ${\phi}_R (\xi)$ as well. The situation
is the same as in the plain-wave quantization scheme in MS where
the boundary condition (\ref{BoundCond}) does not necessarily
follow from the self-adjointness of the operator ${\cal
K}_{M}(z)$.

If one deals with the states $|g\ket$ for which
\begin{equation}
\label{RegCond}
\bra
g|K^{-1}|g\ket=\intR\frac{d\mu}{\mu^2}\,|g(\mu)|^2=
2\intR\frac{d\rho}{\rho}|\psi_g(\rho)|^2<\infty,
\end{equation}
then the boundary condition Eq.(\ref{BC_zero}) may be proved
straightforwardly in the same way as it was done in section
\ref{Minkowski}. The condition Eq.(\ref{RegCond})  corresponds
to the regularity condition which was proposed by Kay in
Ref.\cite{Kay} to guarantee the existence of thermal states for
Fulling quantization system.

At the end of the section let us discuss the asymptotic behaviour
of Whightman function $D^{(+)}(\eta-\eta',\rho,\rho')$ for fixed
value of $\rho'$ and $\rho\to 0$. In this case the two-point
commutator vanishes and the contribution of small $\mu$ to the
integral in Eq.(\ref{WightFuncPlus}) becomes principal. Thus we
have
\begin{equation}
\label{WF_asympt}
D^{(+)}(\eta-\eta',\rho,\rho')=
\frac{iK_0(m\rho')}{\pi\ln(2/m\rho)}+O(\ln^{-2}(1/m\rho)),
\quad (\Delta s^2<0).
\end{equation}
We see that physically significant quantities die out at point
$\rho=0$ in RS as they do at spatial infinity in MS.

\section{Quantization of a neutral scalar field in D=1+1
Minkowski spacetime (boost modes)} \label{BoostModesSec}

The Rindler observer world line coincides with one of the orbits
of Lorentz group and $R$-wedge is one of domains of MS invariant
under Lorentz rotation. Therefore it is convenient dealing with
the Unruh problem  to quantize the field in the basis of
Eigenfunctions of Lorentz boost operator rather than in the
plane-wave basis.

Since the secondary quantized boost operator $L=M_{tx}$ does not
commute with the energy $H$ and momentum $P$ operators,
\begin{equation}
\label{HPL}
[L,H]=iP,\quad [L,P]=iH,
\end{equation}
it is not diagonal in terms of annihilation, creation operators of
particles with given momentum $p$. In order to diagonalize the
operator $L$ we replace first the variable $p$ by the rapidity
$q=\artanh(p/\epsilon_p)$ and introduce new operators $\alpha_{q}$
as
\begin{equation}
\label{a}
\alpha_q=(m\cosh q)^{1/2} a_p,\quad
[\alpha_q,\hc{\alpha_{q'}}]=\delta(q-q').
\end{equation}
The energy, momentum and boost operators are expressed in terms of new
operators as follows
\begin{equation}
\label{EnMom}
H=m\intRR dq\,\cosh{q}\hc{\alpha_q}\alpha_q,
\quad P=m\intRR dq\,\sinh{q}\hc{\alpha_q}\alpha_q,
\quad L=\frac{i}2\intRR dq\,\hc{\alpha_q}\LRD{q}\alpha_q.
\end{equation}
It is easy to see that in terms of operators $b_{\kappa}$ which
are Fourier transforms of operators $\alpha_{q}$
\begin{equation}
\label{b} b_{\kappa}=\frac1{\sqrt{2\pi}} \intRR e^{i\kappa
q}\alpha_q\,dq, \quad \alpha_q=\frac1{\sqrt{2\pi}} \intRR
e^{-i\kappa q}b_{\kappa}\,d\kappa.
\end{equation}
the boost operator $L$ is diagonal
\begin{equation}
\label{L}
L=\intRR \kappa\,\hc{b_{\kappa}}b_{\kappa}\,d\kappa.
\end{equation}

Using the definition (\ref{a}) and the relations (\ref{b}) one can
easily transform the Eq.(\ref{Quant_PW}) to the form
\begin{equation}
\label{Quant_BM}
\phi_M(x)=\intRR d\kappa\{b_{\kappa}\Psi_{\kappa}(x)+
\hc{b_{\kappa}}\Psi_{\kappa}^*(x)\},
\end{equation}
where functions $\Psi_{\kappa}$ are defined by the integral
representation \cite{NR,Gerlach,JETP,PhLet}
\begin{equation}
\label{Boost_Modes}
\Psi_{\kappa}(x)=\frac1{2^{3/2}\pi} \intRR dq\,
\exp\{im(z\sinh{q}-t\cosh{q})-i\kappa q\}.
\end{equation}
It is assumed that an infinitely small negative imaginary part
added to $t$, see Appendix \ref{CALCULATING_B}. It may be shown
that functions (\ref{Boost_Modes}) are Eigenfunctions of the boost
generator ${\cal B}$,
\begin{equation}
\label{B}
{\cal B}\Psi_{\kappa}(x)=\kappa\Psi_{\kappa}(x), \quad
-\infty<\kappa<+\infty,\quad {\cal B}=i(z\diff{t}+t\diff{z}).
\end{equation}
They are orthonormal relative to inner
product in MS,
\begin{equation}
\label{ortho_b}
(\Psi_{\kappa},\Psi_{\kappa'})_M=\delta(\kappa-\kappa'),\quad
(\Psi_{\kappa}^*,\Psi_{\kappa'})_M=0,
\end{equation}
and together with their adjoints $\Psi_{\kappa}^*$ form a complete
set of solutions for KFG equation in MS. We will call this set of
functions boost modes.

The boost modes can serve as a basis for a new quantization scheme
for the field $\phi_M(x)$. Indeed according to Eqs.(\ref{a}),
(\ref{b}) the commutation relations for $b_{\kappa}$ read
\begin{equation}
\label{b_comm}
[b_{\kappa},\hc{b_{\kappa'}}]=\delta(\kappa-\kappa'),\quad
[b_{\kappa},b_{\kappa'}]=[\hc{b_{\kappa}},\hc{b_{\kappa'}}]=0.
\end{equation}
The vacuum state with respect to operators $b_{\kappa}$ which obeys
the condition
\begin{equation}
\label{Vac_B} b_{\kappa}|0_M\ket=0.
\end{equation}
is exactly the usual Minkowski vacuum. This is because the transition
from the operators $\alpha_q$ to $b_{\kappa}$ (\ref{b}) is a unitary
transformation,
\begin{equation}
\label{Unitary}
b_{\kappa}= F\alpha_{-\kappa}\hc{F}, \quad  F\hc{F}=\hc{F}F=1,
\quad F=\exp\left(i\frac{\pi}4 \intRR dq\,
\left\{\partial_q\hc{\alpha_q}\partial_q\alpha_q+
(q^2-1)\hc{\alpha_q}\alpha_q\right\}\right),
\end{equation}
and hence the solutions $\Psi_{\kappa}$ correspond to positive
frequencies relative to global time $t$.  Note that quantization of
scalar field defined by Eqs.(\ref{b_comm}), (\ref{Vac_B}),
(\ref{Quant_BM}) is equivalent to the one performed in
Ref.\cite{Boul} by analytical extension of Green functions.

There exists another representation of boost modes
\cite{JETP,PhLet} (see also Ref.\cite{GMR} for the fermion case)
corresponding to splitting of MS into the right (R), future (F),
left (L) and past (P) wedges, see Fig.\ref{Fig2},
\begin{equation}
\label{RFLP}
\Psi_{\kappa}= \theta(x_+)\theta(-x_-)\Psi_{\kappa}^{(R)}+
\theta(x_+)\theta(x_-)\Psi_{\kappa}^{(F)}+
\theta(-x_+)\theta(x_-)\Psi_{\kappa}^{(L)}+
\theta(-x_+)\theta(-x_-)\Psi_{\kappa}^{(P)},
\end{equation}
where $x_{\pm}=t\pm x$ are null coordinates in MS. By performing
integration in Eq.(\ref{Boost_Modes}) under assumption $x_+>0$,
$x_-<0$ and using integral representation for Macdonald functions
\cite{BE1} we obtain for $\Psi_{\kappa}^{(R)}$
\begin{mathletters}
\begin{equation}
\label{BM_R}
\Psi_{\kappa}^{(R)}=\frac1{\pi\sqrt{2}}\,
\exp\left(\frac{\pi\kappa}2-i\frac{\kappa}2\ln
\left(\frac{x_+}{-x_-}\right)\right)
K_{i\kappa}\left(m\sqrt{-x_-x_+}\right).
\end{equation}
The explicit form for the boost modes in other wedges may be obtained
from (\ref{BM_R}) by analytical extension. The branch points of the
function Eq.(\ref{BM_R}) lie on the light cone and for transition
from one wedge to another one should use the substitutions
$(-x_-)\to x_-e^{i\pi}$ for the transition $R\to F$,
$x_+\to (-x_+)e^{-i\pi}$ for $F\to L$, $x_-\to (-x_-)e^{-i\pi}$
for $L\to P$ and $(-x_+)\to x_+e^{i\pi}$ for $P\to R$.
Thus we obtain
\begin{equation}
\label{BM_F}
\Psi_{\kappa}^{(F)}=\frac{-i}{2^{3/2}}\,
\exp\left(\frac{\pi\kappa}2-i\frac{\kappa}2\ln
\left(\frac{x_+}{x_-}\right)\right)
H_{i\kappa}^{(2)}\left(m\sqrt{x_-x_+}\right).
\end{equation}
\begin{equation}
\label{BM_L}
\Psi_{\kappa}^{(L)}=\frac1{\pi\sqrt{2}}\,
\exp\left(-\frac{\pi\kappa}2-i\frac{\kappa}2\ln
\left(\frac{-x_+}{x_-}\right)\right)
K_{i\kappa}\left(m\sqrt{-x_-x_+}\right).
\end{equation}
\begin{equation}
\label{BM_P}
\Psi_{\kappa}^{(P)}=\frac{i}{2^{3/2}}\,
\exp\left(-\frac{\pi\kappa}2-i\frac{\kappa}2\ln
\left(\frac{-x_+}{-x_-}\right)\right)
H_{i\kappa}^{(1)}\left(m\sqrt{(-x_-)(-x_+)}\right).
\end{equation}
\end{mathletters}
After transition $P\to R$ we return to Eq.(\ref{BM_R}). The second
linearly independent set of solutions for KFG equation
$\Psi_{\kappa}^*$ may be obtained from Eqs.(\ref{BM_R}) -
(\ref{BM_P}) by substitutions $-x_{\pm}\to x_{\pm}e^{-i\pi}$. The
possibility of unique recovery of the values of $\Psi_{\kappa}(x)$
(and hence the values of the field $\phi_M(x)$) in the full MS
using its values only in $R$-wedge and the requirement of
positivity of the energy is an illustration of the content of the
Reeh-Schlieder theorem (see e.g. Refs.\cite{RS,SW}).

The splitting (\ref{RFLP}) corresponds to the four families of
orbits of the two-dimensional Lorentz group (compare to \S 6 of
Chapter V in Ref.\cite{Vil}). As it was already mentioned the
functions Eqs.(\ref{BM_R}) - (\ref{BM_P}) have branch points at
the light cone which corresponds to the four degenerate orbits
$x_{\pm}=0$, ${\rm sgn} t=\pm 1$. For example if $t\to z>0$ then
using the expansion of Macdonald function $K_{\nu}(\zeta)$ for
$\zeta\to 0$ \cite{BE1} we obtain
\begin{mathletters}
\begin{equation}
\label{BM_R_LC}
\Psi_{\kappa}^{(R)}=\frac1{2^{3/2}\pi}\,e^{\pi\kappa/2}
\left\{\Gamma(i\kappa)\left(\frac{mx_+}2\right)^{-i\kappa}+
\Gamma(-i\kappa)\left(-\frac{mx_-}2\right)^{i\kappa}+...\right\}.
\end{equation}
The light cone asymptotic behaviour of the other functions
Eqs.(\ref{BM_F}) - (\ref{BM_P}) may be derived from
Eq.(\ref{BM_R_LC}) by the described above procedure of analytical
extension. For example,
\begin{equation}
\label{BM_F_LC}
\Psi_{\kappa}^{(F)}=\frac1{2^{3/2}\pi}\,e^{\pi\kappa/2}
\left\{\Gamma(i\kappa)\left(\frac{mx_+}2\right)^{-i\kappa}+
e^{-\pi\kappa}\Gamma(-i\kappa)
\left(\frac{mx_-}2\right)^{i\kappa}+...\right\}.
\end{equation}
\end{mathletters}
After substituting these expressions in Eq.(\ref{RFLP}) we find
the light-cone behaviour of the boost mode which reads
\begin{equation}
\label{BM_LC}
\Psi_{\kappa}(x)=\frac1{2^{3/2}\pi}\,e^{\pi\kappa/2}
\left\{\Gamma(i\kappa)\left(\frac{mx_+}2-i0\right)^{-i\kappa}+
\Gamma(-i\kappa)\left(-\frac{mx_-}2+i0\right)^{i\kappa}+...\right\}.
\end{equation}
The distributions $(\zeta\pm
i0)^{\lambda}=\zeta^{\lambda}\theta(\zeta)+ e^{\pm
i\lambda\pi}(-\zeta)^{\lambda}\theta(-\zeta)$ in Eq.(\ref{BM_LC})
were defined and studied in Ref.\cite{GSh}. It is clear from
Eq.(\ref{BM_LC}) that in spite of the presence of
$\theta$-functions in Eq.(\ref{RFLP}) the modes $\Psi_{\kappa}(x)$
obey the KFG equation without sources. In the vertex of the light
cone $t=z=0$ (which is the fixed point for the Lorentz group) from
Eq.(\ref{Boost_Modes}) we get
\begin{equation}
\label{BM_OR}
\Psi_{\kappa}(0,0)=\frac1{\sqrt{2}}\delta(\kappa).
\end{equation}
The same result may be derived either from Eqs.(\ref{BM_R}) -
(\ref{BM_P}) by taking into account that \cite{LSS}
\begin{equation}
\label{K(0)}
\frac{i}2\, H_{i\kappa}^{(1)}(0)= -\frac{i}2\,
H_{i\kappa}^{(2)}(0)= \frac1{\pi}\, K_{i\kappa}(0)=\delta(\kappa),
\end{equation}
or from Eq.(\ref{BM_LC}). This result means that all modes
$\Psi_{\kappa}(x)$ except for the singular zero mode vanish at the
vertex of the light cone.

The expression for the annihilation operator $b_\kappa$ in terms
of field operator on an arbitrary Cauchy surface (compare
Eqs.(\ref{FieldTo_a}),(\ref{c_phi_R})) read
\begin{equation}
\label{b_phi}
b_{\kappa}=(\Psi_{\kappa},\phi_M)_M=
i\intRR\Psi_{\kappa}^*(t,z)\LRD{t} \phi_M(t,z)\,dz,\quad t\ne 0.
\end{equation}
For the surface $t=0$ we have \cite{JETP}
\begin{equation}
\label{b_phi1}
\begin{array}{r}\nl
b_{\kappa}=
\frac{i}{\pi\sqrt{2}}\left( e^{\pi\kappa/2}
\intR F_R(z,\kappa)\,dz+
e^{-\pi\kappa/2}
\int\limits_{-\infty}^0 F_L(z,\kappa)\,dz\right) ,
\quad F_{R,L}=K_{i\kappa}(\pm mz)
\left(\DiffM{\phi_M}{t}\mp\DiffM{\phi_M}{z}\right)_{t=0}\pm\EL\EL
\pm\Gamma(\mp i\kappa)\left(\pm\frac{mz}2\right)^{\pm i\kappa}
\left(\DiffM{\phi_M}{z}\right)_{t=0}+m\left\{
K_{i\kappa\mp 1}(\pm mz)-\frac12\Gamma(1\mp i\kappa)
\left(\pm \frac{mz}2\right)^{\pm i\kappa-1}\right\}\phi_M(0,z),
\end{array}
\end{equation}
with upper (lower) signs corresponding to the indices $R$ ($L$)
consequently. (The derivation of this very important formula is
given in Appendix \ref{CALCULATING_B}).

Note that calculation of the Whightman function
$\Delta^{(+)}(x,m)$ by use of Eqs.(\ref{Quant_BM}), (\ref{Vac_B})
of course leads to the result (\ref{DeltaPlus}). It gives the
independent proof that the plain waves quantization is unitary
equivalent to the boost modes one.


\section{The Unruh construction}
\label{UnruhQuant}

In this Section we will consider the quantum field theory aspect
of the Unruh problem. The study of this problem was inspired by
Fulling who suggested \cite{Fulling} in 1973 a valid scheme for
quantization of a massive scalar field in RS (see
Sec.\ref{Rindler}). Fulling treated RS as a part of MS and hence
considered Fulling-Rindler vacuum as a state of quantum field in
MS. Therefore he tried to express the annihilation and creation
operators of Fulling-Rindler particles (\ref{c_phi_R}) in terms of
plain-wave operators (\ref{FieldTo_a}) and argued that the
Minkowski vacuum state could be interpreted as a many-particle
Fulling-Rindler state. In virtue of the boundary conditions which
the field $\phi_R$ must obey in RS and which we have considered in
Sec.\ref{Rindler} Fulling procedure in MS is physically
meaningless. But even if one disregards the existence of boundary
conditions the quantization scheme suggested by Fulling is
incorrect for MS since it implied the assumption that the field
modes which he used for quantization were equal to zero outside
$R$-wedge of MS. In other words only the first term from
Eq.(\ref{RFLP}) for boost modes was involved in the procedure of
quantization. But due to the presence of $\theta$-functions this
term obeys not the KFG equation for the free field in MS but the
equation with sources of infinite power localized on the light
cone.

To avoid this difficulty Unruh with the help of a rather elegant
trick made an attempt to construct a new scheme of quantization
which should be valid in MS and in some sense repeat the Fulling
scheme in $R$-wedge. The central point of Unruh suggestion
\cite{Unruh} was to use such superpositions $R_{\mu}$, $L_{\mu}$
of boost modes $\Psi_{\kappa}$ with positive and $\Psi_{\kappa}^*$
with negative frequencies that they vanish either in the left or
right wedges of MS and coincide with Fulling modes respectively in
the right or left wedges. We will present the explicit form of the
Unruh modes and discuss the Unruh quantization scheme in
Sec.\ref{SUBUA}. Then in the Sec.\ref{SUBUB} we will discuss the
so-called Unruh effect.

\subsection{The Unruh quantization}
\label{SUBUA}

The Unruh modes can be expressed in terms of boost modes as
follows
\begin{equation}
\label{Unruh_Modes} R_{\mu}(x)=\frac1{\sqrt{2\sinh\pi\mu}}
\left\{e^{\pi\mu/2}\Psi_{\mu}(x)-
e^{-\pi\mu/2}\Psi_{-\mu}^*(x)\right\}, \quad
L_{\mu}(x)=\frac1{\sqrt{2\sinh\pi\mu}}
\left\{e^{\pi\mu/2}\Psi_{-\mu}^*(x)-
e^{-\pi\mu/2}\Psi_{\mu}(x)\right\},
\end{equation}
where $\mu>0$.
These functions obey the normalization conditions
\begin{equation}
\label{UM_norm}
(R_{\mu},R_{\mu'})_M=-(L_{\mu},L_{\mu'})_M=\delta(\mu-\mu'),\quad
(R_{\mu},R_{\mu'}^*)_M=(L_{\mu},L_{\mu'}^*)_M=
(R_{\mu},L_{\mu'})_M=(R_{\mu},L_{\mu'}^*)_M=0.
\end{equation}
With the help of Eqs.(\ref{BM_R}) - (\ref{BM_P}) one can easily
check that for $x$ belonging to the right wedge $R$ the "left"
Unruh modes $L_{\mu}(x)$ vanish while the "right" modes
$R_{\mu}(x)$ coincide with Fulling modes $\Phi_{\mu}(\xi)$,
Eq.(\ref{Full_Modes}). For the light cone behaviour of Unruh modes
in the right wedge $R$ we have
\begin{equation}
\label{UM_R_LC_R}
R_{\mu}(x)=\Phi_{\mu}(\xi)\sim\frac{\sqrt{\sinh\pi\mu}}{2\pi}\,
\left\{\Gamma(i\mu)\left(\frac{mx_+}2\right)^{-i\mu}+
\Gamma(-i\mu)\left(-\frac{mx_-}2\right)^{i\mu}\right\},
\quad x\in R,\quad x_{\pm}\to 0.
\end{equation}
In the future wedge $F$ they can be represented as
\begin{mathletters}
\begin{equation}
\label{UM_R_F}
R_{\mu}(x)=-\frac{i}{2\sqrt{\sinh\pi\mu}}\,
\left(\frac{x_+}{x_-}\right)^{-i\mu/2}
J_{-i\mu}\left(m\sqrt{x_+x_-}\right)\sim
\frac{\sqrt{\sinh\pi\mu}}{2\pi}\,
\Gamma(i\mu)\left(\frac{mx_+}2\right)^{-i\mu},
\quad x\in F,\quad x_{\pm}\to 0,
\end{equation}
\begin{equation}
\label{UM_L_F}
L_{\mu}(x)=\frac{i}{2\sqrt{\sinh\pi\mu}}\,
\left(\frac{x_+}{x_-}\right)^{-i\mu/2}
J_{i\mu}\left(m\sqrt{x_+x_-}\right)\sim
-\frac{\sqrt{\sinh\pi\mu}}{2\pi}\,
\Gamma(-i\mu)\left(\frac{mx_-}2\right)^{i\mu},
\quad x\in F,\quad x_{\pm}\to 0.
\end{equation}
\end{mathletters}

Note that in spite of Unruh statement made in Ref.\cite{Unruh}
functions (\ref{Unruh_Modes}) can not be obtained by analytical
extension of Fulling modes (\ref{Full_Modes}) through the future
and past horizons into the $F$- and $P$-wedges. As a matter of
fact the Unruh modes are combinations of functions defined on
different sheets of Riemannian surface, see the rules for
enclosing of brunch points of the boost modes in
Sec.\ref{BoostModesSec}.

By inverting the relations Eq.(\ref{Unruh_Modes}) and substituting
the result in Eq.(\ref{Quant_BM}) one obtains
\begin{equation}
\label{Quant_UM}
\phi_M(x)=\intR d\mu\,\{r_{\mu}R_{\mu}(x)+\hc{r_{\mu}}R_{\mu}^*(x)+
l_{\mu}L_{\mu}^*(x)+\hc{l_{\mu}}L_{\mu}(x)\}.
\end{equation}
The Eq.(\ref{Quant_UM}) holds everywhere in MS except the origin
of Minkowski coordinate frame because the boost mode
$\Psi_{\kappa}(x)$ is singular at the point $x=(0,0)$ (see
Eq.(\ref{BM_OR})) and hence it is impossible to perform
integration in Eq.(\ref{Quant_BM}) at this point over positive and
negative values of $\kappa$ independently. This means that
expansion (\ref{Quant_UM}) is valid for the field in MS with cut
out point $x=(0,0)$ \footnote{This fact was earlier proposed as
the reason for "thermal properties" of "Minkowski vacuum" in
Refs.\cite{CrDuff,Troost}.}.

The Unruh operators $r_{\mu}$ and $l_{\mu}$ from
Eq.(\ref{Quant_UM}) are defined by the expressions \cite{Unruh,UW}
\begin{equation}
\label{Unruh_Ops}
r_{\mu}=\frac1{\sqrt{2\sinh\pi\mu}}
\left\{e^{\pi\mu/2}b_{\mu}+e^{-\pi\mu/2}\hc{b_{-\mu}}\right\},
\quad l_{\mu}=\frac1{\sqrt{2\sinh\pi\mu}}
\left\{e^{\pi\mu/2}b_{-\mu}+e^{-\pi\mu/2}\hc{b_{\mu}}\right\},
\quad (\mu>0).
\end{equation}
Using Eqs.(\ref{b_comm}) one can easily show that these operators obey
the following commutation relations
\begin{equation}
\label{CCR_U_Ops}
[r_{\mu},\hc{r_{\mu'}}]=[l_{\mu},\hc{l_{\mu'}}]=\delta(\mu-\mu'),
\quad [r_{\mu},r_{\mu'}]=[l_{\mu},l_{\mu'}]=[r_{\mu},l_{\mu'}]=
[r_{\mu},\hc{l_{\mu'}}]=0.
\end{equation}

Though the Unruh operators satisfy canonical commutation relations
(\ref{CCR_U_Ops}) it is not enough they could be considered as
annihilation, creation operators. A necessary condition for the
latter is existence of a stationary ground state for the system,
vacuum state. Neither the Unruh modes (\ref{Unruh_Modes}) nor
their adjoints are positive frequency solutions for KFG equation
in $P$- and $F$-wedges with respect to {\it any} timelike variable
and hence the Unruh operators (\ref{Unruh_Ops}) are composed of
creation and annihilation operators which relate to particles with
opposite sign of frequency. Therefore it is clear that in the
global MS a stationary vacuum state with respect to the $r$- and
$l$-"particles" can not exist.

To confirm this point let us see how the first term in parentheses
in Eq.(\ref{Quant_UM}) behave in $F$-wedge at small $\mu$  (see
Eqs.(\ref{BM_R}), (\ref{Unruh_Modes}), (\ref{Unruh_Ops}))
\begin{equation}
\label{Rr_small_mu}
r_{\mu}R_{\mu}(x)\sim
-\frac{i}{2^{3/2}\pi\mu}J_0\left(m\sqrt{x_+x_-}\right)\,
(b_0+\hc{b_0})\propto\frac1{\mu},\quad \mu\to 0,\quad x\in F.
\end{equation}
It is clear that the contribution of this term to the integral
(\ref{Quant_UM}) diverges logarithmically at the lower limit.
Similarly we have
\begin{equation}
\label{Ll_small_mu}
\hc{l_{\mu}}L_{\mu}(x)\sim
\frac{i}{2^{3/2}\pi\mu}J_0\left(m\sqrt{x_+x_-}\right)\,
(b_0+\hc{b_0})\propto\frac1{\mu},\quad \mu\to 0,\quad x\in F.
\end{equation}
Though singularities cancel in the sum of these two terms in
Eq.(\ref{Quant_UM}) it is clear that none of the matrix elements
of these parts of the field operator $ \phi_M(x)$ exist
separately. The same is true for contributions of terms with all
other Unruh operators. This situation certainly holds also for the
$P$-wedge.

This consideration shows that the Unruh operators
(\ref{Unruh_Ops}) {\it may not be interpreted} as annihilation,
creation operators and the Unruh construction {\it may not be
regarded} as a valid quantization scheme in the global MS.

Nevertheless we may consider the integral (\ref{Quant_UM}) only
for the world points located entirely in $R$- and $L$-wedges i.e.
inside the double Rindler wedge. It is important that being
restricted to the double Rindler wedge the l.h.s. of
Eq.(\ref{Quant_UM}) cannot be considered as the field $\phi_M(x)$
in MS. Indeed taking into account that the Unruh (Eq.
(\ref{Unruh_Modes})) and Fulling (Eq.(\ref{Full_Modes})) modes
coincide in the $R$-wedge we may represent the l.h.s. of
Eq.(\ref{Quant_UM}) there in the form
\begin{equation}
\label{UMinR} \tilde{\phi}_M(x)=\lim\limits_{\varepsilon\to
0}\int\limits_\varepsilon^\infty d\mu\,\{r_{\mu}\Phi_{\mu}(x)+
\hc{r_{\mu}}\Phi_{\mu}^*(x)\},\quad x\in R.
\end{equation}
Then considering for the sake of simplicity the case of equal
times $t=t'=0$ we have
\begin{equation}
\label{DW Wf} \bra
0_M|\tilde{\phi}_M(0,z)\tilde{\phi}_M(0,z')|0_M\ket=\frac{1}{\pi^{2}}
\lim\limits_{\varepsilon\to 0}\int\limits_\varepsilon^\infty
d\mu\, \cosh{\pi\mu} K_{i\mu}(mz) K_{i\mu}(mz'), \quad z,z'> 0.
\end{equation}
Since $(1/2\pi)K_{i\mu}(0)=0$ at $\mu> 0$ (see Eq.(\ref {K(0)}))
we obtain for $z'=0$
\begin{equation}\label{Wfz0}
\bra 0_M|\tilde{\phi}_M(0,z)\tilde{\phi}_M(0,0)|0_M\ket=0.  
\end{equation}
At the same time for the Whightman function in MS we have
\begin{equation}\label{Wfz0MS}
\bra 0_M|\phi_M(0,z)\phi_M(0,0)|0_M\ket=\frac{1}{2\pi^{2}} \intRR
d\kappa\, \cosh{\pi\kappa} K_{i\kappa}(mz)
K_{i\kappa}(0)=\frac{1}{2\pi}K_0(mz), \quad z> 0 , 
\end{equation}
in full agreement with Eq.(\ref{DeltaPlus}). It is clear after
comparison of Eqs.(\ref{Wfz0}),(\ref{Wfz0MS}) that
$\Delta^{+}(x,m)$ is not equal to zero in the $R$-wedge only due
to existence of singular zero boost mode (\ref{BM_OR}). Since this
mode is absent in the Unruh set (\ref{Unruh_Modes}) the latter is
not a complete set of solutions for the KFG equation. Therefore
from this moment we will denote the l.h.s. of Eq.(\ref{Quant_UM})
as $\phi_{DW}(x)$ instead of $\phi_M(x)$.

Consider quantization of the field $\phi_{DW}(x)$ in the double
Rindler wedge now. Since there exists a timelike variable, namely
Rindler time, with respect to which the Unruh modes
(\ref{Unruh_Modes}) are positive frequency solutions of KFG
equation we can attach to the Unruh operators the meaning of
annihilation, creation operators of particles living in the double
Rindler wedge. But the fields in $R$- and $L$-wedges are
absolutely independent of each other since any two points
belonging to different wedges are separated by spacelike interval.
Therefore double Rindler wedge is a disjoint union of $R$- and
$L$-wedges and quantization in these wedges should be carried out
separately. We have discussed already quantization procedure in
the $R$-wedge and found that it implies existence of boundary
condition ensuring finiteness of the field energy. It is clear
that the field in double Rindler wedge $\phi_{DW}(x)$ should
satisfy the same boundary condition. Taking into account these
considerations we should rewrite Eq.(\ref{Quant_UM}) in the form

\begin{equation}
\label{Quant_DW}
\phi_{DW}(x)= \intR d\mu\,\{r_{\mu}R_{\mu}(x)+
\hc{r_{\mu}}R_{\mu}^*(x)\}+\intR d\mu\,
\{l_{\mu}L_{\mu}^*(x)+\hc{{l_{\mu}}}L_{\mu}(x)\},
\quad x\in R\cup L,
\end{equation}
where Unruh operators should coincide with the corresponding
Fulling operators $c_{\mu}$, $\hc{c_{\mu}}$ and
$c_{\mu}'$, $\hc{{c_{\mu}'}}$ for particles living in $R$- and
$L$-wedges respectfully if the field $\phi_{DW}(x)$ satisfies the
boundary condition
\begin{equation}
\label{BC_DW}
\phi_{DW}(0,0)=0.
\end{equation}
We will prove the latter statement for the operator $r_\mu$ as an
example.

Substituting the expression (\ref{b_phi}) for operators $b_\kappa,
b_\kappa^{\dag}$ into the first formula in Eq.(\ref{Unruh_Ops}) we
obtain after transition to Rindler coordinates (\ref{CoordConn})
\begin{mathletters}
\begin{equation}
\label{r_phi1}
r_{\mu}=\lim_{\rho\to 0}(\tilde
c_{\mu}(\rho)-c_{\mu}^{(s)}(\rho)),
\end{equation}
\begin{equation}
\label{tilde_c}
\tilde c_{\mu}(\rho)= \frac{i}{\sqrt{2\mu}}\,
\int\limits_{\rho}^{\infty}\frac{d\rho}{\rho}\,
\psi_{\mu}(\rho)\left(\DiffM{\phi_M}{\eta}
-i\mu\phi_M\right)_{\eta=0},
\end{equation}
\begin{equation}
\label{r1}
c_{\mu}^{(s)}(\rho)=\frac{i\sqrt{\sinh\pi\mu}}{2\pi}\,
\phi_M(0,\rho)\left\{\Gamma(-i\mu)\left(\frac{m\rho}2\right)^{i\mu}-
\Gamma(i\mu)\left(\frac{m\rho}2\right)^{-i\mu}\right\},
\end{equation}
\end{mathletters}
Note that both $\tilde c_{\mu}(\rho)$ and $c_{\mu}^{(s)}(\rho)$
become singular for $\rho\to 0$ if $\phi_M(0,0)\ne 0$ but the
singularities cancel in their difference in Eq.(\ref{r_phi1}). It
is clear from Eqs.(\ref{r_phi1}), (\ref{r1}) that although
Eqs.(\ref{tilde_c}), (\ref{c_phi_R}) look similar, one can not
identify the right Unruh operator $r_{\mu}$ and the Fulling
annihilation operator $c_{\mu}$ unless
\begin{equation}
\label{BC_MS}
\phi_M(0,0)=0.
\end{equation}
This condition should be understood in weak sense of course.

From Eqs.(\ref{Quant_BM}), (\ref{BM_OR}) we however formally have
\begin{equation}
\label{phi_zero}
\phi_M(0,0)=\frac{b_0+\hc{b_0}}{\sqrt{2}}=
\intRR\frac{dp}{\sqrt{4\pi\epsilon_p}}\,(a_p+\hc{a_p}),
\end{equation}
and therefore for the value of one-particle amplitude (\ref{OPA})
at the vertex of the light cone we obtain
\begin{equation}
\label{OPA_zero}
\phi_f(0,0)=\intRR\frac {dp}{\sqrt{4\pi\epsilon_p}}\,f(p).
\end{equation}
Of course there are no any physical reasons to require vanishing of
this quantity in MS. Thus if one understands Eq.(\ref{BC_MS}) as a
condition for the field in MS it means nothing but cutting out the
point $t=z=0$. Cutting out even a single point is not however a
"painless operation" for MS because it dramatically changes its
properties. In particular MS looses its property to be a globally
hyperbolic spacetime. As a consequence the Cauchy data
(\ref{props_comm}) for the two-point commutator corresponds now to
zero solution of KFG equation unlike what we have in real MS.
Since in four-dimensional MS the Pauli-Jordan function has Cauchy
data similar to (\ref{props_comm}) this result is not a specific
property of two dimensional case.

This inconsistency disappears if we apply the Unruh construction
to the double Rindler wedge rather than to MS. That means that we
should use the expansion (\ref{Quant_DW}) instead of
(\ref{Quant_UM}), substitute $\phi_{DW}$ instead of $\phi_M$ in
Eqs.(\ref{tilde_c}),(\ref{r1}) and read Eq.(\ref{BC_MS}) as
Eq.(\ref{BC_DW}). As a result we conclude that the Unruh
construction is a valid quantization scheme only in the double
Rindler wedge.

\subsection{The Unruh "effect"}
\label{SUBUB}

There are several ways to "give proof" of existence of the Unruh
effect in the frames of conventional quantum field theory
\cite{Unruh,Scia,BD,GMR,Tak,GF,GMM,Wald}.

One of them is based on the equation
\begin{equation}
\label{rr_average}
\bra 0_M|\hc{r_{\mu}}r_{\mu'}|0_M\ket =
\left(e^{2\pi\mu}-1\right)^{-1}\,\delta(\mu-\mu'),
\end{equation}
which can be easily obtained using Eqs.(\ref{Unruh_Ops}),
(\ref{Vac_B}). The l.h.s. of the Eq.(\ref{rr_average}) at
$\mu = \mu'$ is interpreted after integration over $\mu$ as
Minkowski vacuum expectation value of the operator of Fulling
particles, while the r.h.s. of this equation after the standard
trick is written as
\[ \delta(\mu-\mu')\vert_{\mu=\mu'}= \intRR
{e^{i(\mu-\mu')\eta}}\vert_{\mu=\mu'}\,\frac{d\eta}{2\pi}
=\frac{g\Delta\tau}{2\pi}, \]
where $\tau=\eta/g$ is proper time
and $g$ is proper acceleration of Rindler observer. Then
Eq.(\ref{rr_average}) is transformed to the form
\begin{equation}
\label{Unruh1}
\frac{\Delta \bar
N}{\Delta\tau}=\intR\frac{d\omega}{2\pi}\,
(e^{2\pi\omega/g}-1)^{-1},
\end{equation}
where $\omega=g\mu$ is commonly understood as the energy of
Fulling quanta. Finally the integrand in Eq.(\ref{Unruh1}) is
identified with the thermal spectrum corresponding to Davies-Unruh
temperature (\ref{DUT}).

However, as it was shown in the previous section, one can not
identify Unruh operators $r_{\mu}$, $\hc{r_{\mu}}$ with Fulling
annihilation and creation operators $c_{\mu}$, $\hc{c_{\mu}}$ in
MS. Moreover operators $r_{\mu}$, $\hc{r_{\mu}}$ can not serve as
annihilation and creation operators of {\it any} particles in MS.
Therefore the l.h.s. of Eq.(\ref{rr_average}) can not be
interpreted as Minkowski vacuum expectation value of number of
particles.

The Unruh operators coincide with the corresponding Fulling
operators for the field obeying the boundary condition
(\ref{BC_DW}) in the double Rindler wedge. But an observer living
in RS can not define Minkowski vacuum state. To conclude that the
state of the field is Minkowski vacuum one must have possibilities
to perform measurements in every point of Cauchy surface in the
whole MS. This is impossible for an observer living in $R$- (or
$L$-) wedge because he can not perform measurements at the points
belonging to $L$- (or $R$-) wedge. From mathematical point of view
this statement is a direct consequence of Reeh-Schlieder theorem,
see Refs.\cite{RS,SW}. On the other hand observers living in the
double Rindler wedge are not able to perform such measurements due
to existence of the boundary condition (\ref{BC_DW}).

Let us also note that the Bose factor in the r.h.s. of
Eq.(\ref{rr_average}) should not be necessarily interpreted as
thermal distribution. This factor appears entirely due to specific
properties of Bogolubov transformation (\ref{Unruh_Ops}) and is
encountered in many physical problems where in no way does the
notion of temperature arise. Two-mode squeezed photon states in
quantum optics \cite{BarK,JOpt} is a well known example of such
situation, see also Ref.\cite{Klysh}. Two-dimensional harmonic
oscillator can serve as another example, see Appendix
\ref{inequiv}.

Another "derivation" of the Unruh effect is based on the relation
\cite{Unruh,Wald}
\begin{equation}
\label{mvaccont}
|0_M\ket=Z^{-\frac12}\sum\limits_{n=0}^{\infty}
\int\limits_{0}^{\infty}d\mu_{1}\ldots\int\limits_{0}^{\infty}
d\mu_{n}\,e^{-\pi\sum\limits_{i=1}^{n}\mu_i}
|1_{\mu_1},\ldots1_{\mu_n}\ket_{L}\otimes
|1_{\mu_1},\ldots1_{\mu_n}\ket_{R}.
\end{equation}
This formula "determines $r$- and $l$-particle content of
Minkowski vacuum" and allows one to introduce the density matrix
describing states of the field in $R$-wedge, see e.g.
\cite{Scia,Sew,Tak,GF,GMM,Wald}. The latter is achieved by taking
the tensor product of the r.h.s. of Eq.(\ref{mvaccont}) with its
dual and then taking trace over the states of the field in the
$L$-wedge. So for an arbitrary observable ${\cal R}$ depending on
the "values of the field" $\phi_M(x)$ for $x$ belonging only to
the right Rindler wedge $R$ we have
\begin{equation}
\label{main1}
\bra 0_M|{\cal R}|0_M\ket={\rm Sp}\,({\cal R}\rho_R),
\quad \rho_R=Z^{-1}\exp(-H_R/T_{DU}).
\end{equation}
In this equation $\rho_R$ is the density matrix and
$H_R=gK=g\int_0^{\infty}\mu\hc{c_{\mu}}c_{\mu}\,d\mu$
is a secondly quantized Hamiltonian with respect to proper time
$\tau$ of the accelerating observer.

However the l.h.s. of Eq.(\ref{mvaccont}) may not be considered as
Minkowski vacuum state because as we have shown in the previous
section the notion of Rindler particles which is essentially used
in derivation of Eq.(\ref{mvaccont}) makes sense only in double
Rindler wedge rather than in global MS. Therefore
Eq.(\ref{mvaccont}) could describe vacuum state only in double
Rindler wedge. But it loses any physical meaning if one takes into
account the existence of boundary condition (\ref{BC_DW}). Indeed
the derivation of Eq.(\ref{mvaccont}) assumes (see e.g.
\cite{Wald}) that one-particle Hilbert space in the double Rindler
wedge is a direct sum of one-particle Hilbert spaces in $R$- and
$L$-wedges ${\cal H}_{DW}={\cal H}_R\oplus{\cal H}_L$ and Fock
space of states of the field in double Rindler wedge ${\cal
F}({\cal H}_{DW})$ is a tensor product of Fock spaces on ${\cal
H}_R$ and ${\cal H}_L$, ${\cal F}({\cal H}_{DW})\cong{\cal
F}({\cal H}_R)\otimes{\cal F}({\cal H}_L)$. But in virtue of
boundary condition, which is equivalent to cutting out the vertex
of light cone, $R$- and $L$-wedges have no common points and
therefore never interact. Therefore only such superpositions of
state vectors from ${\cal F}({\cal H}_{DW})$ can have physical
sense which do not contain correlations between $r$- and
$l$-particles. This "superselection rule" prohibits states of the
type Eq.(\ref{mvaccont}).

Besides the normalization constant $Z$ in Eq.(\ref{mvaccont})
which also has the meaning of partition function in
Eq.(\ref{main1}) is infinite, namely
\begin{equation}
\label{Z}
Z=\exp(\delta(0)\pi^2/12),
\end{equation}
see Appendix \ref{inequiv}. The divergence of constant $Z$ means
that representations of the canonical commutation relations in
terms of Unruh and boost modes operators are unitary inequivalent.

There could exist two ways to formulate Eq.(\ref{main1}) in
mathematically meaningful way and avoid these difficulties. The
first one is to place the field in a box which in this problem
could be constructed by two uniformly accelerated mirrors moving
in right and left Rindler wedges \cite{CR}. However such
regularization again leads to consideration of double RS as a
physical spacetime of the observer. The second opportunity is to
use algebraic approach and a notion of KMS state as a definition
of thermal equilibrium state.  We discuss below the formulation of
Eq.(\ref{main1}) which was developed in the frame of algebraic
approach to quantum field theory.


\parindent 0.7cm
\section{Algebraic approach}
\label{AlgebraicAppr}

Algebraic approach to quantum theory \cite{Emch,WK,Haag,Wald}
allows one to compare states which cannot be represented by
vectors or density matrices in the same Hilbert space
representation of algebra of observables of the system. It is
because the states in this approach are primarily considered as
positive normalized linear functionals over the algebra of
observables rather than vectors in Hilbert space. The physical
meaning of the state $\W$ in algebraic approach is that the value
$\W(A)$ is the expectation value of the observable $A$ in the
state $\W$. The algebraic counterpart of usual thermal equilibrium
state is called the KMS state \cite{HHW,FR,Haag}. Unlike usual
thermal equilibrium state the KMS state exists even if partition
function of the system diverges. On the language of algebraic
approach the Unruh effect means that the algebraic state
corresponding to Minkowski vacuum state coincides with the KMS
state for double Fulling quantization. In this section we will
show that such conclusion implies existence of boundary condition
at the origin of Minkowski reference frame. Our consideration will
make clear that in algebraic derivation of the Unruh effect the
same inconsistencies are present as in traditional approach.

\subsection{One mode model}

In order to use simple and suitable notation, let us first present
the construction of KMS state over the one mode quantum system
(i.e. one-dimensional harmonic oscillator). We will see that the
case of free Bose field in $D=1+1$ MS requires just a trivial
generalization.

Let $R$ be one-mode quantum system. Its algebra of observables
(which we denote by $\Alg_R$) may be characterized either in terms
of unbounded generators, annihilation and creation operators $r$,
$\hc{r}$ satisfying commutation relation
\begin{equation}
\label{cr}
[r,\hc{r}]=1,
\end{equation}
or in terms of Weyl generators
\begin{equation}
\label{wg}
W(f)=\exp(f r-f^*\hc{r}),
\end{equation}
labeled by complex number $f$ \footnote{The unbounded operators
$r$, $\hc{r}$ may be expressed in terms of Weyl generators, for
example $r=\half (W'(f)-iW'(if)) \vert_{f=0}$ where derivatives
are taken with respect to $f$.} and which satisfy the following
requirements
\begin{equation}
\label{wr}
W(f_1)W(f_2)=\exp(\half(f_1^* f_2-f_2^* f_1))
W(f_1+f_2),\quad W(f)^*=W(-f).
\end{equation}
An arbitrary observable $\Robs$ may be written in the form
$\Robs=\Robs(r,\hc{r})$. Time evolution of observables is determined
by the equation
\begin{equation}
\label{te}
\Robs(t)=\hc{U}(t)\Robs U(t),\quad U(t)=\exp(-iH_R t),
\end{equation}
with $H_R=\eps \hc{r} r$ being the one-mode Hamiltonian. In
particular,
\begin{equation}
\label{ptcr}
\Robs(t)=\Robs(r e^{-i\eps t},\hc{r} e^{i\eps t}),
\quad W(f,t)=W(f(t)),\quad f(t)=fe^{-i\eps t}.
\end{equation}
Vacuum state $\W_R$ is defined by the relation $r|0_R\ket=0$ and the
Hilbert space $\HH_R$ where the considered operators act is
generated by the basis $|n_R\ket=(\hc{r})^n/\sqrt{n!} |0_R\ket$.
The algebraic vacuum state $\W_R$ is a prescription for
calculating expectation values in the vacuum state
\begin{equation}
\label{vac}
\W_R(\Robs)=\bra 0_R|\Robs(r,\hc{r})|0_R\ket.
\end{equation}
The vacuum expectation value of Weyl generator $W(f)$ may be
easily shown to be equal to
\begin{equation}
\label{vexv}
\W_R(W(f))=\exp(-\half |f|^2).
\end{equation}
Thermal equilibrium state with inverse temperature $\beta$ is
defined as follows
\begin{equation}
\label{therm}
\W_R^{(\beta)}(\Robs)={\rm Sp}\,(\Robs\rho_{\beta}),\quad
\rho_{\beta}=Z_{\beta}^{-1}\exp(-\beta H_R)=
Z_{\beta}^{-1}\sum\limits_{n}\exp(-\beta\eps n) |n_R\ket\bra n_R|,
\end{equation}
where $Z_{\beta}=\sum\limits_{n}\exp(-\beta\eps n)$. Of course
$Z_{\beta}$ is finite for this simple one-mode model,
$Z_{\beta}=(1-e^{-\beta\eps})^{-1}$. But since this may be not the
case for quantum systems with infinite number of degrees of
freedom it is important to reformulate Eq.(\ref{therm}) in the
form not containing the value of $Z_{\beta}$ explicitly.

In order to do it one should introduce another copy of system $R$,
say $L$ which does not interact with $R$, and consider the
combined quantum system $R\otimes L^*$ , the "double system".
\footnote{Normally in applications of KMS states to statistical
mechanics this additional copy of initial system is considered as
a mathematical trick used with the purpose to describe thermal
state by a vector in Hilbert space and there are no attempts to
interpret it as a really existent.} The asterisk here indicates
that we choose Hamiltonian of the combined system to be
\begin{equation}
\label{NewH}
H_{R\otimes L^{*}}=H_R\otimes 1-1\otimes H_L
\end{equation}
rather than $H_R\otimes 1+1\otimes H_L$ (compare to Eq.(\ref{Osc})).
One can interpret it saying that time direction at $L$ is inverted.
The vacuum state of the system $R\otimes{L}^*$ is defined by the
relations
\[  r|0_{R\otimes{L}^*}\ket=0,\quad l|0_{R\otimes{L}^*}\ket=0,  \]
where $l$ is annihilation operator for $L$ and the vectors
\[ |n_R\ket\otimes |m_L\ket= \frac{(\hc{r})^n}{\sqrt{n!}}\,
\frac{(\hc{l})^m}{\sqrt{m!}}\, |0_{R\otimes{L}^*}\ket, \]
constitute the basis of the Hilbert space $\HH_{R\otimes{L}^*}$.
Let us introduce the state
\begin{equation}
\label{Omega}
|\Omega_{\beta}\ket=Z_{\beta}^{-1/2}\sum\limits_{n}\exp(-\beta\eps
n/2) |n_R\ket\otimes |n_L\ket.
\end{equation}
It can be immediately verified that the thermal expectation value
(\ref{therm}) may be rewritten in the form
\begin{equation}
\label{therm1}
\W_R^{(\beta)}(\Robs)=\bra\Omega_{\beta}|\Robs(r\otimes 1,
\hc{r}\otimes 1)|\Omega_{\beta}\ket,
\end{equation}
where calculation is performed in the space $\HH_{R\otimes{L}^*}$.
Note that the expectation value (\ref{therm1}) does not depend on
time. This property served as a reason for the choice of
Hamiltonian $H_{R\otimes{L}^*}$ in the form (\ref{NewH}).

Now let us consider operators
\begin{equation}
\label{bb}
b_{+}=\cosh\theta\,r\otimes 1-\sinh\theta\,1\otimes\hc{l}, \quad
b_{-}=-\sinh\theta\,\hc{r}\otimes 1+\cosh\theta\,1\otimes l,
\end{equation}
where $\theta$ is defined by the equation
$\tanh\theta=e^{-\beta\eps/2}$. Note that these operators depend
on time as $b_{\pm}(t)\propto e^{\mp i\eps t}$. The key
observation is that these operators annihilate the state
$|\Omega_{\beta}\ket$ and together with $\hc{b_{+}}$, $\hc{
b_{-}}$ satisfy the usual commutation relations
\begin{equation}
\label{b_props} b_{\pm}|\Omega_{\beta}\ket=0, \quad
[b_{+},b_{-}]=0, \quad [b_{+},\hc{b_{-}}]=0, \quad
[b_{\pm},\hc{b_{\pm}}]=1.
\end{equation}
The span of the vectors of the form
\[ \frac{(\hc{b_{+}})^n}{\sqrt{n!}}\,
\frac{(\hc{b_{-}})^m}{\sqrt{m!}}\, |\Omega_{\beta}\ket \]
constitutes the Hilbert space $\HH$ which in our one-mode case
coincides with the space $\HH_{R\otimes{L}^*}$. Expressing
operators $r, \hc{r}$ in terms of operators $b_{\pm}$ we can
rewrite Eq.(\ref{therm1}) in the form
\begin{equation}
\label{therm2} \W_R^{(\beta)}(\Robs)=\bra\Omega_{\beta}|
\Robs(b_{+}\cosh\theta+\hc{b_{-}}\sinh\theta,\,
b_{-}\sinh\theta+\hc{b_{+}}\cosh\theta)|\Omega_{\beta}\ket.
\end{equation}
The r.h.s. of this equation does not contain $Z_{\beta}$ and in
virtue of Eqs.(\ref{b_props}) is just the vacuum expectation value
of the observable $\Robs$ but calculated in the space $\HH$ and
with respect to vacuum $|\Omega_{\beta}\ket$. In general the
algebraic state defined by Eq.(\ref{therm2}) is called "the KMS
state". \footnote{Another equivalent definition of KMS state is
given by the requirement $\W_R^{(\beta)}(\Robs_1(t)\Robs_2(t'))=
\W_R^{(\beta)}(\Robs_2(t')\Robs_1(t+i\beta))$ where $\Robs_1$,
$\Robs_2$ are arbitrary observables of system $R$,
\cite{Kubo,MS,HHW,Umez}.}

In our case the KMS state (\ref{therm2}) is just the usual thermal
equilibrium state. One can generalize the definition of KMS state
to the observables of the double system $R\otimes {L}^*$ which
have the form $A(r,\hc{r},l,\hc{l})$ by setting
\begin{equation}
\label{therm3} \W^{(\beta)}(A)=\bra\Omega_{\beta}|
A(b_{+}\cosh\theta+\hc{b_{-}}\sinh\theta,\,
b_{-}\sinh\theta+\hc{b_{+}}\cosh\theta,\,
b_{-}\cosh\theta+\hc{b_{+}}\sinh\theta,\,
\hc{b_{-}}\cosh\theta+b_{+}\sinh\theta) |\Omega_{\beta}\ket.
\end{equation}
The state defined by Eq.(\ref{therm3}) is called the double KMS state
\cite{Kay}. Given definitions of KMS and double KMS states in the
evident way may be generalized to the case of any finite or
infinite number of degrees of freedom. For the latter case the
usual definition of thermal equilibrium state is in general not
valid.

Let us give the formulas for expectation values of Weyl generators
in KMS and double KMS states. By simple computation one gets
\begin{equation}
\label{kms1}
\W_R^{(\beta)}(W(f))=\exp\left(
-\frac12\coth\left(\frac{\beta\eps}2\right)
|f|^2\right).
\end{equation}
We define the Weyl generator $W(f_R,f_L)$ for the double system by
the equation
\begin{equation}
\label{weyl1}
W(f_R,f_L)= \exp(f_R r-f_R^* \hc{r}-f_L^* l+ f_L\hc{l}).
\end{equation}
The advantage of this definition is that time dependence of Weyl
generator takes the form $W(f_R,f_L,t)=W(f_R e^{-i\eps t},f_L
e^{-i\eps t})$ or in other words both $f_R$ and $f_L$ are positive
frequency solutions for "one-mode" wave equation
$(-\partial_t^2+\eps^2)f=0$. The expectation value of the operator
(\ref{weyl1}) in the double KMS state may be shown to be equal to
\begin{equation}
\label{dkms1}
\W^{(\beta)}(W(f_R,f_L))=\exp\left\{
-\frac12\coth\left(\frac{\beta\eps}2\right)(|f_R|^2+|f_L|^2)
-\frac1{\sinh(\beta\eps/2)}{\rm Re}\,f_R^*f_L
\right\}.
\end{equation}

\subsection{The Unruh problem in algebraic approach}

Now let us turn back to the Unruh problem. At first sight
Eqs.(\ref{bb}) and the definition of the state
$|\Omega_{\beta}\ket$ (\ref{b_props}) look very similar to
inverted Eqs.(\ref{Unruh_Ops}) expressing boost operators
$b_{\kappa}$ in terms of Unruh operators $r_{\mu}$, $l_{\mu}$ and
the definition of the state $|\Omega_M\ket$ in Eq.(\ref{Vac_B}).
But we will show that it is not correct to apply the notion of
double KMS state to the Unruh problem. The physical reason is that
free field in Minkowski spacetime cannot be decomposed into two
non-interacting fields living in the interior of right and left
Rindler wedges.

To reformulate the Eq.(\ref{main1}) in terms of algebraic approach
let us introduce the required definitions. The algebra $\Alg$ of
observables of the free field in MS is a $C^*$ algebra with Weyl
generators $W(\Phi)=\exp\{-(\phi_M,\Phi)_M\}$ where $\phi_M(x)$ is
the quantum field operator and $\Phi(x)$ is a real-valued solution
of KFG equation (\ref{KFG_M}). More precisely $\Alg$ contains
arbitrary finite linear combinations of Weyl generators and their
limits in the sense of convergence in $C^*$ norm. For a complete
set of positive frequency orthonormal modes
$\Upsilon_{\lambda}(x)$ we define $f_{\lambda}$ by the equation
$f_{\lambda}=(\Upsilon_{\lambda},\Phi)_M$ and rewrite $W(\Phi)$ in
the form
\begin{equation}
\label{weyl3}
W(\Phi)=\exp\left\{\int\limits d\lambda\,
(f_{\lambda}\Ao_{\lambda}-f_{\lambda}^*
\hc{\Ao_{\lambda}})\right\},
\end{equation}
where $\Ao_{\lambda}$ is the appropriate annihilation operator.
The Weyl relations take form
\begin{equation}
\label{CCR}
W(\Phi_1)W(\Phi_2)=\exp\left\{\frac12 \int\limits
d\lambda\,(f_{1\lambda}^*f_{2\lambda}-
f_{2\lambda}^*f_{1\lambda})\right\} W(\Phi_1+\Phi_2),\quad
W(\Phi)^*=W(-\Phi),
\end{equation}
(compare to Eqs.(\ref{wg}),(\ref{wr})). Note that solutions $\Phi(x)$
are required to decrease sufficiently fast at spatial infinity (say,
have compact support on any Cauchy surface).

The expectation value of Weyl generator $W(\Phi)$ in Minkowski
vacuum state $\W_M$ may be obtained by generalization of the
formula (\ref{vexv}):
\begin{equation}
\label{MV}
\W_M(W(\Phi))=\exp\left(-\frac12
\intRR d\kappa |f_{\kappa}|^2\right),
\end{equation}
where coefficients $f_{\kappa}=(\Psi_{\kappa},\Phi)_M$ are defined
with respect to a complete set of boost modes $\Psi_{\kappa}$
(\ref{Boost_Modes}). By inverting relations (\ref{Unruh_Modes}) one
can rewrite eq.(\ref{MV}) in terms of Unruh modes. The result is
\begin{equation}
\label{MV1} \W_M(W(\Phi))=\exp\left\{-\frac12 \intR d\mu
\left(\coth\pi\mu\,(|f_{\mu}^{(L)}|^2+|f_{\mu}^{(R)}|^2)+
\frac2{\sinh\pi\mu}{\rm Re}\,(f_{\mu}^{(L)})^*f_{\mu}^{(R)}
\right)\right\},
\end{equation}
where $f_{\mu}^{(R)}=(R_{\mu},\Phi)_M$, $f_{\mu}^{(L)}=(L_{\mu},\Phi)_M$.

Finite linear combinations of elements from $\Alg$ of the form
$W(\Phi)$ with $\Phi$ vanishing in the closed wedge $\bar L$ and
the limits of sequences of such linear combinations in uniform
sense (i.e. limits in the sense of convergence in $C^*$ norm)
constitute a $C^*$ subalgebra $\RWA$ of $\Alg$ which is called the
right wedge algebra. The left wedge algebra $\LWA$ and the double
wedge algebra $\DWA$ are defined similarly by restricting to
solutions which vanish in closed wedge $\bar R$ and in a
neighborhood of $h_0$ respectively, see Fig.\ref{Fig2}.

Now let us evaluate expectation value of Weyl generator in a
double KMS state with temperature $\beta^{-1}$ with respect to
Fulling quantization prescription. By generalizing
Eq.(\ref{dkms1}) one obtains \footnote{Compare to section 1.4 in
Ref.\cite{Kay}.}
\begin{equation}
\label{MV2} \tilde\W_F^{(\beta)}(W(\Phi))=\exp\left\{-\frac12\intR
d\mu\, \left(\coth\left( \frac{\beta\mu}2\right)\,
(|\zeta_{\mu}^{(R)}|^2+|\zeta_{\mu}^{(L)}|^2)
+\frac2{\sinh(\beta\mu/2)}{\rm Re}\,
(\varphi_{\mu}^{(L)})^*\zeta_{\mu}^{(R)} \right)\right\},
\end{equation}
where $\Phi=\Phi_R\oplus\Phi_L$,
$\zeta_{\mu}^{(R)}=(\Phi_{\mu}^{(R)},\Phi)_R$,
$\zeta_{\mu}^{(L)}=(\Phi_{\mu}^{(L)},\Phi)_L$, $\Phi_{\mu}^{(R)}$
is complete set of Fulling modes (\ref{Full_Modes}) and
$\Phi_{\mu}^{(L)}$ is their analog in the wedge $L$. Note that the
expression in the r.h.s. of Eq.(\ref{MV2}) is well-defined only if
test functions $\Phi$ in double RS obey the requirement
\[\intR\frac{d\mu}{\mu}\left\vert\zeta_{\mu}^{(R,L)}
\right\vert^2<\infty,\] which is referred as regularity condition
in Ref.\cite{Kay} (compare to Eq.(\ref{RegCond})).


Let us obtain the relation between the coefficients
$f_{\mu}^{(R)}$ and $\zeta_{\mu}^{(R)}$ in Eqs.(\ref{MV1}),
(\ref{MV2}). For this purpose we first evaluate
$f_{\mu}^{(R)}=(R_{\mu},\Phi)_M$ supposing that the surface of
integration is a surface of constant positive small $t$ and then
take limit $t\to 0$. For Unruh mode we use the expression
\begin{equation}
\label{UM_spl}
R_{\mu}(x)=R_{\mu}^{({\rm R})}(x)\theta(x_+)\theta(-x_-)+
R_{\mu}^{({\rm F})}(x)\theta(x_+)\theta(x_-)+R_{\mu}^{({\rm P})}(x)
\theta(-x_+)\theta(-x_-),
\end{equation}
which can be easily obtained from Eqs.(\ref{RFLP},\ref{Unruh_Modes}).
To calculate the inner product
\begin{equation}
\label{f_mu}
f_{\mu}^{(R)}=i\intRR dz\,R_{\mu}^*(x)\LRD{t}\Phi(x),
\end{equation}
we need a time derivative of (\ref{UM_spl}). Taking into account
that $t>0$ we write it in the following way:
\begin{equation}
\label{TDer}
\begin{array}{r}\nl
\frac{\partial}{\partial t}R_{\mu}(x)=
\left(\frac{\partial}{\partial t}R^{({\rm R})}_{\mu}(x)\right)
\theta(x_+)\theta(-x_-)+\left(\frac{\partial}{\partial
t}R^{({\rm F})}_{\mu}(x)\right) \theta(x_+)\theta(x_-)+
R^{({\rm F})}_{\mu}(x)\delta(x_+)+ \EL
+\left\{R^{({\rm F})}_{\mu}(x)-R^{({\rm R})}_{\mu}(x)\right\}
\delta(-x_-).
\end{array}
\end{equation}
It is not very hard to verify using Eqs.(\ref{Unruh_Modes}),
(\ref{Full_Modes}), (\ref{BM_R}) and (\ref{CoordConn}) that
\begin{equation}
\label{First} \lim_{t\to 0} \, i\intR dz\, R_{\mu}^{({\rm
R})*}(x)\LRD{t}\Phi(x)=\zeta_{\mu}^{(R)}.
\end{equation}
The second term in the r.h.s. of Eq.(\ref{TDer}) vanishes when
$t\to 0$. Therefore we consider only the last two terms.
Substituting the expansions Eq.(\ref{UM_R_LC_R}) and
Eq.(\ref{UM_R_F}) into Eq.(\ref{f_mu}) and taking limit $t\to 0$
we obtain \footnote{Compare to Eqs.(\ref{r_phi1})-(\ref{r1}), see
also Appendix \ref{CALCULATING_B}.}
\begin{equation}
\label{rel} f_{\mu}^{(R)}=\zeta_{\mu}^{(R)}+
\frac{i}{2\pi}\sqrt{\sinh\pi\mu}\,\lim_{z\to 0}\Phi(0,z)
\left\{\Gamma(i\mu)\left(\frac{mz}2\right)^{-i\mu}-
\Gamma(-i\mu)\left(\frac{mz}2\right)^{i\mu}\right\},
\end{equation}
and the similar relation between $f_{\mu}^{(L)}$ and
$\zeta_{\mu}^{(L)}$.

One concludes after comparing Eqs.(\ref{MV1}),(\ref{MV2}) that
equation
\begin{equation}
\label{main}
\W_M(W(\Phi))=\tilde\W_F^{(2\pi)}(W(\Phi)),
\end{equation}
holds if and only if
\begin{equation}\label{BC AA}
\Phi(0,0)=0,
\end{equation}
(compare to Eq.(\ref{BC_DW}) and hence by linearity
\begin{equation}
\label{Unruh}
\W_M=\tilde\W_F^{(2\pi)}\quad \mbox{on $\DWA$}.
\end{equation}
This equation is an analog of Eq.(\ref{main1}) in algebraic
approach (see Ref.\cite{Kay}). The restriction of Eq.(\ref{Unruh})
to the right wedge algebra $\RWA$ is usually referred as
Bisogniano - Wichmann theorem \cite{BW,BW1,Kay,Haag}.

We see that Eq.(\ref{Unruh}) holds only on the double wedge
subalgebra $\DWA\subset \Alg$, which corresponds to the space of
those solutions for the field equation which satisfy the boundary
condition (\ref{BC AA}).  The r.h.s. of Eq.(\ref{Unruh}) doesn't
admit continuation to the whole algebra $\Alg$ while the l.h.s.
admits such continuation. Therefore functionals $\W_M$ and
$\tilde\W_F^{(2\pi)}$ describe \tit{different} algebraic states
over the algebra of observables of the free field in MS.

Let us consider two opportunities to interpret Eq.(\ref{Unruh}).
The first one is to treat $\Alg$ as the true algebra of
observables for the accelerated observer. In this case
Eq.(\ref{Unruh}) does not hold for all observables and therefore
Minkowski vacuum does not coincide with the thermal state
$\tilde\W_F^{(2\pi)}$.

The second opportunity is to propose that $\DWA$ should be the
true algebra of observables for accelerated observer. In this case
the true Minkowski vacuum state $\W_M$ (which is the state over
the algebra $\Alg$) is unrealizable state for such observer. Then
Eq.(\ref{Unruh}) is satisfied for all physical observables and
hence the {\it restriction} $\W_M\vert_{\DWA}$ of the state $\W_M$
to $\DWA$ coincides with the state $\tilde\W_F^{(2\pi)}$ and
admits interpretation in terms of Fulling -- Unruh quanta. But let
us stress that the Minkowski vacuum state is physically
distinguished among the other possible states of the theory not so
much due to it's explicit expression (which of course is inherited
by it's restrictions to the subalgebras of $\Alg$) as by it's key
physical properties such as Poincar\'e invariance, spectral
conditions, local commutativity and cluster property (see the
Whightman reconstruction theorem, \cite{SW}). Although some of
these properties are inherited by the restrictions to subalgebras,
the other properties such as Poincar\'e invariance generally are
not \footnote{Lacking of Poincar\'e invariance for field theory
with boost time evolution is a consequence of the fact that
1-parameter boost group does not constitute normal subgroup in
Poincar\'e group.}. But exactly these properties are mentioned
when one assumes that some quantum system is prepared initially in
the state of Minkowski vacuum. Since subalgebra $\DWA$ is not
Poincar\'e invariant there are no any physical reasons to consider
the state $\W_{M}\vert_{\DWA}$ as the initial state of the field.
Moreover, since the automorphisms of $\DWA$ corresponding to boost
time evolution don't mix up observables from $\RWA$ and $\LWA$
there is still no way for Rindler observer to prepare the state
$\W_{M}\vert_{\DWA}$. We see that consideration of Unruh problem
in algebraic approach leads to the same results as in usual
field-theoretical approach.


\section{Conclusions}
\label{Conclusions}

We have analyzed the Unruh problem in the frame of quantum field
theory and have shown that the Unruh quantization scheme is valid
in the double Rindler wedge rather than in MS. The double Rindler
wedge is composed of two disjoint regions which causally do not
communicate with each other. Moreover the Unruh construction
implies existence of boundary condition at the point $h_0$ (or
2-dimensional plain in the case of $1+3$-dimensional spacetime) of
MS. Such boundary condition may be interpreted as a topological
obstacle which gives rise to a superselection rule prohibiting any
correlations between $r$- and $l$-particles. Thus a Rindler
observer living in the $R$-wedge in no way can influence the part
of the field from the $L$-wedge and therefore elimination of the
invisible "left" degrees of freedom will take no effect for him.
Hence averaging over states of the field in one wedge can not lead
to thermalization of the state in the other.

In algebraic approach the Unruh effect is commonly identified with
the Bisogniano - Wichmann theorem. According to the Bisogniano -
Wichmann theorem the Minkowski vacuum expectation value of only
those observables which are entirely localized in the interior of
the $R$-wedge constitutes the algebraic state which satisfies the
KMS condition with respect to Rindler timelike variable $\eta$.
This statement implies two points essential for its physical
interpretation. First, it is assumed that the observer which
carries out measurements lives in MS. Only then he could prepare
the Minkowski vacuum state as the initial state of the field.
Second, the variable $\eta$ must coincide with proper time of the
observer. Only then he can interpret the KMS state as a thermal
bath with Davies - Unruh temperature. But the Rindler observer can
carry out measurements only inside the $R$-wedge and hence can not
prepare the Minkowski vacuum state. From the other hand the
variable $\eta$ can not coincide with proper time of an observer
which is an inertial one at least asymptotically in far past and
far future. Nevertheless only such observer for whom inertial in-
and out- regions exist is able to prepare a state with finite
number of particles in MS. These are the reasons why the
Bisogniano - Wichmann theorem is irrelevant for consideration of
the Unruh problem.

Hence considerations of the Unruh problem both in the standard and
algebraic formulations of quantum field theory lead us to
conclusion that principles of quantum field theory does not give
any grounds for existence of the "Unruh effect".

Nevertheless there exists another aspect of the Unruh problem
dealing with behaviour of a particular detector uniformly
accelerated in MS. The direct consideration of the behavior of a
constantly accelerating physical detector is a very difficult
problem and its treatment in literature is very contradictory and
often is simply erroneous. The major difficulty is that an object
moving with a constant proper acceleration must be considered as a
point object. This is because different points of a finite size
body rigid with respect to Rindler coordinate frame move actually
with different accelerations. Thus one should use an elementary
particle or a microscopic bound system as a detector. In both
cases the detector is a \tit{quantum} object moving along a
definite  \tit{ classical} trajectory. Such assumption is in
contradiction with the uncertainty principle, its range of
applicability is very limited and therefore it must be used with
proper care. Moreover, it was shown by Nikishov and Ritus \cite
{NR} that elementary particles placed in a constant electric field
\tit{do not demonstrate} the universal thermal response.

It is clear that a heavy atom for which WKB approach is valid
satisfies physical claims for the  detector much better than an
elementary particle. Unfortunately a systematic relativistic
theory of bound states is still absent. Utilization of non
relativistic bound systems as detectors was discussed in
Ref.\cite{MKP}. The ionization rate of a heavy ion moving with a
constant acceleration was considered. It was shown that the
ionization rate differs from the one obtained by virtue of the
detailed balance principle applied to an atom immersed in a
thermal bath with the Davies-Unruh temperature. It was also shown
that the time of "thermal ionization" (if it was at all possible)
is parametrically much greater than the time of destruction of the
atom due to the tunneling ionization process in electric field.

It is worth to add that in literature the Unruh effect  is usually
explained by existence of event horizons for a constantly
accelerated observer. But we understand that the notion of a
constantly accelerated observer is an inadmissible idealization.
It is clear that for any physical object the horizons are absent.

We certainly understand that behavior of accelerated detectors
will differ from those at rest. We admit that under some
circumstances detectors of some special configuration will follow
Unruh behavior. But no conclusive proof exists that this behavior
is universal and does not depend on the nature of the detector and
the accelerating field.

\acknowledgements

We would like to thank A.A. Starobinski, U. Gerlach and B.S. Kay
for very helpful discussions. N.B. Narozhny and A.M. Fedotov are
grateful to R. Ruffini for hospitality at Rome University "La
Sapienza" (Italy). V.A. Belinskii thanks the Institut des Hautes
Etudes Scientifiques at Bures -- sur -- Yvette (France) where a
part of the work for this paper was done for hospitality and
support. This work was supported in part by the Russian Fund for
Fundamental Research under projects 97--02--16973 and
98--02--17007.

\vspace{1cm}
\appendix
\section{Derivation of Eq.\lowercase{(\ref{b_phi1})}
for {\lowercase{$b_{\kappa}$}}} \label{CALCULATING_B}

To derive Eq.(\ref{b_phi1}) let us introduce small parameter $z_0$
and split the integral in Eq.(\ref{b_phi}) as follows,
\begin{equation}
\label{b_RL}
\begin{array}{l}\nl
b_{\kappa}=b_{\kappa}^{(R)}(z_0)+b_{\kappa}^{(0)}(z_0)
+b_{\kappa}^{(L)}(z_0), \quad
b_{\kappa}^{(R)}=i\int\limits_{z_0}^{\infty} dz\,
{\Psi_{\kappa}^{(R)}}^*(0,z)\left\{\DiffM{\phi_M}{t}
-i\frac{\kappa}{z}\phi_M\right\}_{t=0}, \EL\EL
b_{\kappa}^{(0)}=i\int\limits_{-z_0}^{z_0} dz\,
\left(\Psi_{\kappa}^*(t,z)\LRD{t}\phi_M(t,z)\right)_{t=0}, \quad
b_{\kappa}^{(L)}=i\int\limits_{-\infty}^{-z_0} dz\,
{\Psi_{\kappa}^{(L)}}^*(0,z)\left\{\DiffM{\phi_M}{t}
-i\frac{\kappa}{z}\phi_M\right\}_{t=0}.
\end{array}
\end{equation}
In the first and last integrals of Eq.(\ref{b_RL}) we used
Eq.(\ref{B}) for calculation of the time derivative of boost mode
at $t=0$.

Let us first calculate the first term in Eq.(\ref{b_RL}). Using
Eq.(\ref{BM_R}) for $\Psi_{\kappa}^{(R)}$ we find \footnote{One
should add a small negative imaginary part to $t$ in order to
choose the right branches of the functions contained in the
expression for $\Psi_{\kappa}^{(R)}$.}
\begin{equation}
\label{b_R0}
b_{\kappa}^{(R)}=\frac{i e^{\pi\kappa/2}}{\pi\sqrt{2}}\,
\int\limits_{z_0}^{\infty} dz\, K_{i\kappa}(mz)\left\{
\DiffM{\phi_M}{t}-i\frac{\kappa}{z}\phi_M\right\}_{t=0}.
\end{equation}
Taking into account the well-known formula \cite{BE1}
\[ -\frac{i\kappa}{z}\, K_{i\kappa}(mz)=m\,\{K_{i\kappa}'(mz)+
K_{i\kappa-1}(mz)\}, \]
we may rewrite Eq.(\ref{b_R0}) in the form
\begin{equation}
\label{b_R1}
b_{\kappa}^{(R)}=\frac{i e^{\pi\kappa/2}}{\pi\sqrt{2}}\,
\int\limits_{z_0}^{\infty} dz\,\left\{
K_{i\kappa}(mz)\DiffM{\phi_M}{t}+
m K_{i\kappa}'(mz)\phi_M+
m K_{i\kappa-1}(mz)\phi_M\vphantom{\frac12}\right\}_{t=0}.
\end{equation}
Finally after substitution
$m K_{i\kappa}'(mz)\phi_M=(K_{i\kappa}(mz)\phi_M)_z'-
K_{i\kappa}(mz)(\phi_M)_z'$
and adding and subtracting to the integrand the term
$\Gamma(-i\kappa)\{(mz/2)^{i\kappa}\phi_M\}_z'$
we obtain
\begin{equation}
\label{b_R2}
b_{\kappa}^{(R)}=\frac{i e^{\pi\kappa/2}}{\pi\sqrt{2}}\,
\left\{\int\limits_{z_0}^{\infty} F_R(z,\kappa)\, dz+
\frac12\left(\Gamma(-i\kappa) \vphantom{\frac12}
\left(\frac{mz_0}{2}\right)^{i\kappa}-
\Gamma(i\kappa)\left(\frac{mz_0}{2}\right)^{-i\kappa}\right)
\phi_M(0,z_0)\right\},
\end{equation}
where $F_R$ was defined in Eq.(\ref{b_phi1}). Note that we
have chosen the regularization term by the requirement that
the integral in Eq.(\ref{b_R2}) converges when $z_0$ tends
to zero. Note also that we assumed that the field vanishes
at spatial infinity.

Substitution of Eq.(\ref{BM_L}) into the third integral in
Eq.(\ref{b_RL}) yields
\begin{equation}
\label{b_L0}
b_{\kappa}^{(L)}=\frac{i e^{-\pi\kappa/2}}{\pi\sqrt{2}}\,
\int\limits_{-\infty}^{-z_0} dz\, K_{i\kappa}(-mz)\left\{
\DiffM{\phi_M}{t}-i\frac{\kappa}{z}\phi_M\right\}_{t=0}.
\end{equation}
It is easy to see that the r.h.s. of Eq.(\ref{b_L0}) may be
obtained from the r.h.s. of Eq.(\ref{b_R0}) by changing the
variable of integration $z\to -z$ and substitutions
$\kappa\to -\kappa$, $\phi_M(t,z)\to\phi_M(t,-z)$. Thus we
obtain
\begin{equation}
\label{b_L2}
b_{\kappa}^{(L)}=\frac{i e^{-\pi\kappa/2}}{\pi\sqrt{2}}\,
\left\{\int\limits_{z_0}^{\infty} F_L(-z,\kappa)\, dz+
\frac12\left(\Gamma(i\kappa) \vphantom{\frac12}
\left(\frac{mz_0}{2}\right)^{-i\kappa}-
\Gamma(-i\kappa)\left(\frac{mz_0}{2}\right)^{i\kappa}\right)
\phi_M(0,-z_0)\right\}.
\end{equation}

Since $b_{\kappa}^{(R)}(z_0)+b_{\kappa}^{(L)}(z_0)$ becomes
singular when $z_0$ tends to zero one should also consider the
contribution of the second integral in Eq.(\ref{b_RL}). Using the
integral representation of boost modes (\ref{Boost_Modes}) this
integral may be written in the form
\begin{equation}
\label{b_00}
b_{\kappa}^{(0)}=\frac{i}{2^{3/2}\pi}\int\limits_{-z_0}^{z_0}
dz\intRR dq\,\exp(-imz\sinh{q}+i\kappa q)
\left\{\DiffM{\phi_M}{t}-im\cosh(q)\,\phi_M\right\}_{t=0}.
\end{equation}
Let $z_0$  be sufficiently small we could change $\phi_M(0,z)$ to
$\phi_M(0,0)$. Then after performing integration over $z$ and
changing variable of integration $q\to u=mz_0\sinh{q}$ this
expression may be reduced to
\begin{equation}
\label{b_01}
b_{\kappa}^{(0)}=\frac{\sqrt{2}}{\pi}\phi_M(0,0) \intR
\frac{du}{u}\,\sin{u}\cos(\kappa q(u)),
\end{equation}
(the term with time derivative of the field vanishes when
$z_0$ tends to zero). Since the effective values of the variable
$u$ in the integral (\ref{b_01}) are of order $1$ we may use
approximation
\begin{equation}
\label{cos}
\cos{\kappa q}\approx\frac12\left\{
\left(\frac{mz_0}{2u}\right)^{-i\kappa}
+\left(\frac{mz_0}{2u}\right)^{i\kappa}\right\},
\end{equation}
which is valid if $u\sim 1$ and $z_0$ is small enough. Substitution of
Eq.(\ref{cos}) into Eq.(\ref{b_01}) and evaluation of the integral
yields
\begin{equation}
\label{b_02}
b_{\kappa}^{(0)}=\frac{i\sinh(\pi\kappa/2)}{\pi\sqrt{2}}\phi_M(0,0)\,
\left\{\Gamma(i\kappa)\left(\frac{mz_0}{2}\right)^{-i\kappa}-
\Gamma(-i\kappa)\left(\frac{mz_0}{2}\right)^{i\kappa}\right\}.
\end{equation}
Finally substituting Eqs.(\ref{b_R2}), (\ref{b_L2}) and
(\ref{b_02}) into Eq.(\ref{b_RL}) and taking limit $z_0\to 0$ we
obtain Eq.(\ref{b_phi1}).

\section{Analogy between the Unruh states and squeezed states of a
harmonic oscillator}
\label{inequiv}

\secnum=0
\newsec
Consider  a two dimensional harmonic oscillator in
$\{x,y\}$-plain. The Hamiltonian of such oscillator reads
\begin{equation}
\label{Osc} H_{osc}=\hc{b_{+}}b_{+}+\hc{b_{-}}b_{-}+1,
\end{equation}
where
\[ b_{+}=\frac1{\sqrt{2}}\left(x+\Diff{x}\right),\quad
b_{-}=\frac1{\sqrt{2}}\left(y+\Diff{y}\right), \]
and the commutation relations are of the usual form,
\begin{equation}
\label{ComRels}
[b_{\pm},\hc{b_{\pm}}]=1,\quad
[b_{\pm},\hc{b_{\mp}}]=0.
\end{equation}
The ground state $|0\ket$ satisfies the condition
\begin{equation}
\label{Vac}
b_{\pm}|0\ket=0.
\end{equation}

Let us introduce the new operators $r_{\nu}$, $l_{\nu}$ as
\begin{equation}
\label{rl_Ops} r_{\nu}=S(\nu)b_{+}\hc{S(\nu)}=\cosh\theta\, b_{+}+
\sinh\theta\,\hc{b_{-}}, \quad
l_{\nu}=S(\nu)b_{-}\hc{S(\nu)}=\sinh\theta\, \hc{b_{+}}+
\cosh\theta\, b_{-},
\end{equation}
where the unitary operator $S(\nu)$ reads
\begin{equation}
\label{TRANS}
\begin{array}{c}\nl
S(\nu)=e^{i\theta{\cal G}}=
\exp\{\theta(b_{+}b_{-}-\hc{b_{+}}\hc{b_{-}})\}=
\exp(-e^{-\nu}\hc{b_{+}}\hc{b_{-}})
\exp\{\half\ln(1-e^{-2\nu})H_{osc}\} \exp(e^{-\nu}b_{+}b_{-}),
\EL\EL \nu=-\ln\tanh\theta>0,\quad 0\le\theta<\infty.
\end{array}
\end{equation}
The operators (\ref{rl_Ops}) obey the commutation relations in the
form
\begin{equation}
\label{ComRels1}
[r_{\nu},\hc{r_{\nu}}]=[l_{\nu},\hc{l_{\nu}}]=1,
\quad [r_{\nu},l_{\nu}]=0.
\end{equation}
Note that the generator of the transformation (\ref{TRANS}),
\begin{equation}
\label{G} {\cal G}=-i\left(x\Diff{y}+y\Diff{x}\right),
\end{equation}
looks very similar to the boost operator ${\cal B}$, compare
Eq.(\ref{B}).

According to Eqs.(\ref{ComRels}), (\ref{rl_Ops}) we have
\begin{equation}
\label{rl_Vac}
r_{\nu}|s_{\nu}\ket=0,\quad
l_{\nu}|s_{\nu}\ket=0,
\end{equation}
where
\begin{equation}
\label{rl_Vac1}
|s_{\nu}\ket=S(\nu)|0\ket=(1-e^{-2\nu})^{1/2}
\exp(-e^{-\nu}\hc{b_{+}}\hc{b_{-}})|0\ket,
\end{equation}
is a two dimensional squeezed vacuum state which in contrast to
the ground state $|0\ket$ is not stationary,
\begin{equation}
\label{TIME}
|s_{\nu},t\ket=e^{-iHt}|s_{\nu}\ket=(1-e^{-2\nu})^{1/2}
\exp(-e^{-\nu-2it}\hc{b_{+}}\hc{b_{-}})|0\ket.
\end{equation}
For the amplitude of transition between the ground and squeezed
vacuum states according to Eq.(\ref{rl_Vac1}) we have
\begin{equation}
\label{inter}
\bra 0|s_{\nu}\ket=Z_{\nu}^{-1/2},\quad
Z_{\nu}=(1-e^{-2\nu})^{-1},
\end{equation}
and the expectation value of the "number of squeezed quanta"
$N_r=\hc{r_{\nu}}r_{\nu}$ in the ground state reads
\begin{equation}
\label{Number}
\bra 0|N_{r}|0\ket=(e^{2\nu}-1)^{-1}=
{\rm Sp}\,(N_{r}\rho_{r}),
\end{equation}
where
\begin{equation}
\label{RHO}
\rho_{r}=Z_{\nu}^{-1}\exp(-2\nu N_{r}),
\end{equation}
and the "partition function" $Z_{\nu}$ is defined in
Eq.(\ref{inter}).

The analogy between Eqs.(\ref{inter},\ref{Number}) and
Eq.(\ref{main1}) becomes evident after substitution $\nu=\pi\mu$.
It is clear that the appearance of the Bose factor in
Eq.(\ref{rr_average}) results completely from the properties of
Bogolubov transformation and does not relate to any sort of
thermal behaviour.

\newsec
We will show now that due to infinite number of degrees of freedom
in the Unruh problem the considered analogy is not complete and
representation of canonical commutation relations in terms of the
Unruh operators is unitary inequivalent to the standard one (in
terms of the plain waves or boost operators).

One can write the relation between the Unruh and boost operators
in the form
\begin{equation}
\label{rlb}
r_{\mu}=\cosh\theta_{\mu} b_{\mu}+
\sinh\theta_{\mu} \hc{b_{-\mu}},\quad
l_{\mu}=\sinh\theta_{\mu}\hc{b_{\mu}}+
\cosh\theta_{\mu} b_{-\mu},
\end{equation}
with $\theta_{\mu}$ defined by $\tanh\theta_{\mu}=\exp(-\pi\mu)$.
The vacuum state of the field in the double Rindler wedge
$|0_{DW}\ket$ ( called sometimes the Fulling vacuum, \cite{GMM})
is defined by the requirements
\begin{equation}
\label{Unruh_Vac}
r_{\mu}|0_{DW}\ket=0,\quad l_{\mu}|0_{DW}\ket=0,\quad \mu>0.
\end{equation}
If the Fulling vacuum could be represented by the vector
$|0_{DW}\ket$ in the same Hilbert space where Minkowski vacuum
state $|0_M\ket$ is defined then there should exist a unitary
operator $S$ such that
\begin{equation}
\label{S}
|0_{DW}\ket=S|0_M\ket,\quad Sb_{\mu}\hc{S}=r_{\mu},\quad
Sb_{-\mu}\hc{S}=l_{\mu}.
\end{equation}
A simple calculation shows that such operator $S$ has the
following formal representation (compare to Eq.(\ref{TRANS})):
\begin{equation}
\label{S1}
S=\exp\left(\intR d\mu\theta_{\mu}(b_{\mu}b_{-\mu}-
\hc{b_{\mu}}\hc{b_{-\mu}})\right).
\end{equation}
It is obvious from this representation that $|0_{DW}\ket$ has
the form
\begin{equation}
\label{vU}
|0_{DW}\ket=\left( K^{(0)}+\intR d\mu
K^{(2)}(\mu)\hc{b_{\mu}}\hc{b_{-\mu}}+
\intR d\mu_1\intR d\mu_2 K^{(4)}(\mu_1,\mu_2)
\hc{b_{\mu_1}}\hc{b_{-\mu_1}}\hc{b_{\mu_2}}
\hc{b_{-\mu_2}}+...\right) |0_M\ket.
\end{equation}
Consider the matrix element
\begin{equation}
\label{f}
f[\theta_{\mu}]=\bra 0_{DW}|0_M\ket=
\bra 0_M|\hc{S}[\theta_{\mu}]|0_M\ket,
\end{equation}
for arbitrary function $\theta_{\mu}$. The derivative of the
functional $f[\theta_{\mu}]$ can be expressed as follows (see
e.g. Sec. 2.4 of Ref.\cite{Umez}),
\begin{equation}
\label{de1}
\begin{array}{c}\nl
\frac{\delta f}{\delta\theta_{\mu}}=
-\bra 0_M|b_{\mu}b_{-\mu}\hc{S}|0_M\ket=
\bra 0_M|\hc{S}(Sb_{\mu}\hc{S})(Sb_{-\mu}\hc{S})|0_M\ket=
-\bra 0_M|\hc{S}r_{\mu}l_{\mu}|0_M\ket=\EL
=-f\cosh\theta_{\mu}\sinh\theta_{\mu}\delta(0)-\sinh^2\theta_{\mu}
\frac{\delta f}{\delta\theta_{\mu}}.
\end{array}
\end{equation}
For the last transformation in Eq.(\ref{de1}) we have used
Eqs.(\ref{rlb}) and the obvious formula
\begin{equation}
\label{de2}
\frac{\delta f}{\delta\theta_{\mu}}=
\bra 0_M|\hc{S}\hc{b_{\mu}}\hc{b_{-\mu}}|0_M\ket.
\end{equation}
After simplification of Eq.(\ref{de1}) one obtains for the
functional $f[\theta_{\mu}]$ the differential equation ,
\begin{equation}
\label{de3}
\frac{\delta f[\theta_{\mu}]}{\delta\theta_{\mu}}=
-\delta(0)\tanh\theta_{\mu} f[\theta_{\mu}],
\end{equation}
the formal solution of which we can write as
\begin{equation}
\label{sde}
f=\exp\left(-\delta(0)\intR
d\mu\ln\cosh\theta_{\mu}\right),
\end{equation}
(the constant of integration is fixed by the requirement that
$f$ should be equal to 1 for $\theta_{\mu}=0$). Evaluation of
the integral yields
\[ \intR d\mu\ln\cosh\theta_{\mu}=-\frac12\intR d\mu
\ln(1-e^{-2\pi\mu})= \frac{\pi^2}{24}. \]
Thus we obtain
\begin{equation}
\label{f2}
f=\bra 0_{DW}|0_M\ket=\exp(-\delta(0)\pi^2/24)=0,
\end{equation}
i.e. $K^{(0)}$ in Eq.(\ref{vU}) vanishes. Further with the help of
Eqs.(\ref{de2}),(\ref{sde}),(\ref{f2}) we get
\begin{equation}
\label{f3}
\bra 0_{DW}| \hc{b_{\mu}}\hc{b_{-\mu}}|0_M\ket=
\frac{\delta f[\theta_{\mu}]}{\delta\theta_{\mu}}=
-\delta(0)\exp(-\pi\mu-\delta(0)\pi^2/24)=0,
\end{equation}
i.e. $K^{(2)}=0$. Processing further in such a way we conclude
that $|0_{DW}\ket=0$. It means that there is no Unruh vacuum state
in the same Hilbert space where Minkowski vacuum exists and that
the Unruh operators (\ref{rlb}) form unitary inequivalent
representation of commutation relations.

It is clear from Eq.(\ref{Omega}) (where for current consideration
one should change $|\Omega_{\beta}\ket$ to $|0_M\ket$, $\beta$ to
$2\pi$ and $|0_R\ket\otimes|0_L\ket$ to $|0_{DW}\ket$) that $\bra
0_{DW}|0_M\ket=Z_{2\pi}^{-1/2}$. Therefore we can also express
Eq.(\ref{f2}) in the form
$Z_{2\pi}=\exp(\delta(0)\pi^2/12)=\infty$ (compare to
Eq.(\ref{Z})).

\newpage
\centerline{\uppercase{Figures}}
\vspace{5cm}
\begin{figure}[h]
\caption{Splitting of MS into wedges $R$, $F$, $L$, $P$.}
\label{Fig2}
\end{figure}
%
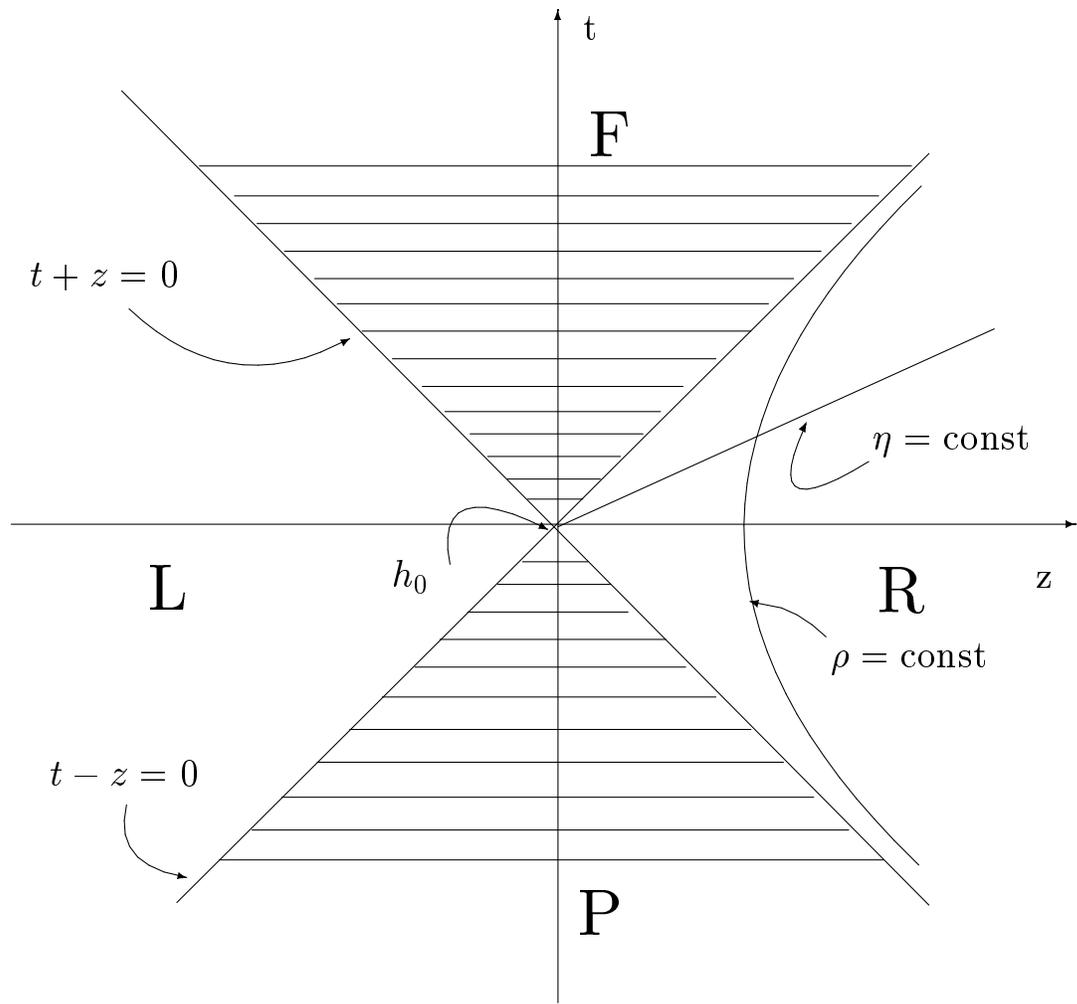
\end{document}